\documentclass[draftclsnofoot,onecolumn,12pt]{IEEEtran}
\usepackage[latin9]{inputenc}
\usepackage{float}
\usepackage{enumitem}
\usepackage{amsmath}
\usepackage{amsthm}
\usepackage{amssymb}
\usepackage{amsthm}
\usepackage{subfig}
\makeatletter

\floatstyle{ruled}
\newfloat{algorithm}{tbp}{loa}
\providecommand{\algorithmname}{Algorithm}
\floatname{algorithm}{\protect\algorithmname}

\theoremstyle{plain}
\newtheorem{thm}{\protect\theoremname}
\theoremstyle{plain}
\newtheorem{prop}[thm]{\protect\propositionname}
\theoremstyle{remark}
\newtheorem{rem}{Remark}
\normalsize
\usepackage{bbm}
\usepackage{adjustbox}
\usepackage{enumitem}
\usepackage{cite}
\usepackage{algorithm}
\usepackage{algorithmic}
\usepackage{amsthm}
\usepackage{empheq}\usepackage{mhsetup}
\usepackage{amsfonts}\usepackage{balance}
\usepackage{epstopdf}
\usepackage{enumitem}
\usepackage[dvipsnames]{xcolor}
\usepackage{pgfplots}
\usepackage{pgfplotstable}
\usepackage{tikz}
\usepackage{graphicx}

\newcommand{\lw}{1.5pt} 

\tikzset{mark options={line width=1pt,solid}}%

\definecolor{myred}{rgb}{0.95,0.00000,0}%
\definecolor{mygray}{rgb}{0.9,0.9,0.9}%
\definecolor{mygreen}{rgb}{0,0.5,0}%
\definecolor{myblue}{rgb}{0.4,0.42,0.8}%
\definecolor{myblack}{rgb}{0.2,0.2,0.2}%
\definecolor{mypurple}{rgb}{0.45,0,0.9}%
\definecolor{mywhite}{rgb}{1,1,1}%
\definecolor{mymagenta}{rgb}{0.95,0,0.95}%

\usepackage{color}

\floatstyle{ruled}
\newfloat{algorithm}{tbp}{loa}
\providecommand{\algorithmname}{Algorithm}
\floatname{algorithm}{\protect\algorithmname}

\IEEEaftertitletext{\vspace{-2\baselineskip}}

\DeclareMathOperator{\EEE}{\mathbb{E}}
\DeclareMathOperator{\var}{\mathbb{V}\mathrm{ar}}

\DeclareMathOperator{\F}{\mathcal{F}}

\DeclareMathOperator{\K}{\mathcal{K}}

\DeclareMathOperator{\E}{\mathbf{E}}

\DeclareMathOperator{\z}{\mathbf{z}}

\DeclareMathOperator{\Hh}{\mathbf{h}}
\DeclareMathOperator{\HH}{\mathbf{H}}

\DeclareMathOperator{\rrr}{\mathbf{r}}

\DeclareMathOperator{\G}{\mathbf{G}}
\DeclareMathOperator{\D}{\mathbf{D}}

\DeclareMathOperator{\LL}{\mathcal{L}}

\DeclareMathOperator{\CN}{\mathcal{CN}}

\DeclareMathOperator{\x}{\mathbf{x}}

\DeclareMathOperator{\y}{\mathbf{y}}
\DeclareMathOperator{\s}{\mathbf{s}}

\DeclareMathOperator{\n}{\mathbf{n}}
\DeclareMathOperator{\U}{\mathbf{U}}

\DeclareMathOperator{\uu}{\mathbf{u}}

\DeclareMathOperator{\g}{\mathbf{g}}

\DeclareMathOperator{\Z}{\mathbf{Z}}

\DeclareMathOperator{\SIgma}{\boldsymbol{\sigma}}

\DeclareMathOperator{\ETA}{\boldsymbol{\eta}}

\DeclareMathOperator{\Epsilon}{\boldsymbol{\epsilon}}

\DeclareMathOperator{\PSI}{{\psi}}

\DeclareMathOperator{\THETA}{{\theta}}

\DeclareMathOperator{\ZETA}{\boldsymbol{\zeta}}

\makeatother

\providecommand{\propositionname}{Proposition}
\providecommand{\theoremname}{Theorem}
\allowdisplaybreaks
\begin{document}
\title{{\Huge{}{}Massive MIMO for Serving Federated Learning and Non-Federated Learning Users}}
\author{Muhammad~Farooq, \IEEEmembership{Graduate Student Member, IEEE},
Tung~Thanh~Vu, \IEEEmembership{Member, IEEE}, Hien~Quoc~Ngo,
\IEEEmembership{Senior Member, IEEE}, and~Le-Nam~Tran, \IEEEmembership{Senior Member, IEEE}
\thanks{Muhammad~Farooq and Le-Nam~Tran are with the School of Electrical
and Electronic Engineering, University College Dublin, Ireland (e-mail:
muhammad.farooq@ucdconnect.ie; nam.tran@ucd.ie).} \thanks{Tung~Thanh~Vu and Hien~Quoc~Ngo is with the Institute of Electronics,
Communications and Information Technology, Queen's University Belfast,
Belfast BT3 9DT, U.K. (email: t.vu@qub.ac.uk; hien.ngo@qub.ac.uk).}}
\maketitle
\begin{abstract}
With its privacy preservation and communication efficiency, federated learning (FL) has emerged as a promising learning framework for beyond 5G wireless networks. It is anticipated that future wireless networks will jointly serve both FL and downlink non-FL user groups in the same time-frequency resource. While in the downlink of each FL iteration, both groups jointly receive data from the base station in the same time-frequency resource, the uplink of each FL iteration requires bidirectional communication to support uplink transmission for FL users and downlink transmission  for non-FL users. To overcome this challenge, we present half-duplex (HD) and full-duplex (FD) communication schemes to serve both groups. More specifically, we adopt the massive multiple-input multiple-output technology and aim to maximize the minimum effective rate of non-FL users under a quality of service (QoS) latency constraint for FL users. Since the formulated problem is highly nonconvex, we propose a power control algorithm based on successive convex approximation to find a stationary solution. Numerical results show that the proposed solutions perform significantly better than the considered baselines schemes. Moreover, the FD-based scheme outperforms the HD-based scheme in scenarios where  the self-interference is small or moderate and/or the size of FL model updates is large.
\end{abstract}

\section{Introduction}

\label{sec:Introd}
The use of mobile phones and wearable devices enables continuous collection and transfer of data \cite{Ometov21SD,Giannetsos11}, which has been the main driving force behind the explosive increase in data mobile traffic in recent years.  Also, due to a constant  growing interest in new features and tools, the computational power of these devices is increasing day by day.  Thus, in many applications, part of data processing is carried out at user's wireless devices. In this context, questions over the transmission of private information over wireless networks naturally arise. To preserve data privacy, a potential solution is to store the data on local servers and move network computation to the edge \cite{li20SPM,Ahmed17Edge}.
In fact, data privacy has drawn significant interest in developing new machine learning techniques that can ensure data privacy and exploit the computational resources of users at the same time. One such a promising technique is knowns as \emph{Federated Learning} (FL) which was first introduced in \cite{mcmahan17AISTATS}. FL is a decentralized form of machine learning  that allows edge devices to learn from a shared prediction model and to keep the data samples on device without exchanging them. Due to this data privacy attribute, FL has been used in a wide range of real-world digital applications e.g., Gboard, FedVision, functional MRI, FedHealth, etc. \cite{Abdulrehman21survery,Aledhari20FL,Chen20FedHealth}.

FL has also gained growing attention from the  wireless communications research community recently due to its privacy protection and resource utilization features \cite{li20SPM,Lim20FLsurvey,chen21TWC,amiri20TWC,Yang2021EEFL,vu20TWC,tran19INFOCOM}, mainly from the viewpoint of implementing FL over wireless networks. 
These pioneer studies can be classified as ``learning-oriented" or
``communication-oriented". The learning-oriented category aims to improve the learning performance (e.g., training loss, test accuracy) subject to inherent factors in wireless networks such as thermal noise, fading, and estimation errors \cite{chen21TWC,amiri20TWC}. Specifically, in \cite{chen21TWC}, Chen \emph{et al.} considered user selection to minimize the FL training loss function under the presence of network constraints. Amiri \emph{et al.} in \cite{amiri20TWC} optimized the test accuracy to schedule devices and allocate power across time slots. The communication-oriented category, on the other hand, focuses on enhancing the communication performance (e.g., training latency, energy efficiency) in the framework of FL \cite{Yang2021EEFL,vu20TWC,tran19INFOCOM}. For example, in \cite{Yang2021EEFL}, Yang \emph{et al.} considered the problem of minimization of the total energy consumption  to train the FL model under a latency constraint. Vu \emph{et al.} in \cite{vu20TWC} focused on minimizing the training latency under transmit power and data rate constraints. In \cite{tran19INFOCOM}, Tran \emph{et al.} investigated the problem of optimizing the computation and communication latencies of mobile devices subject to various trade-offs between the energy consumption, learning time, and learning accuracy parameters. All above-mentioned works take into account serving only FL users (UEs). However, it is certain that future wireless networks will need to serve both the FL and non-FL UEs and thus if FL is to be realized, which calls for novel communication designs. We address this fundamental problem in this paper.

The main challenge in jointly serving FL and non-FL UEs is that it requires  both the uplink and downlink transmission between UEs and the central server occur simultaneously, which has not yet been studied in the existing literature. To understand this, let us briefly describe  a communication round of an FL iteration in the presence of only FL UEs, which consists of four steps: (i) The central server transmits the global update of an ML model to FL UEs; (ii) FL UEs calculate their local model updates based on their local data set; (iii) The local model updates are sent back to the central server; and (iv) The central server calculates the global update by aggregating the received local model updates \cite{vu21SPAWC}. It is clear that problems arise when there are   non-FL  UEs that need to be served in the downlink. First and most importantly,
in Step (iii), the base station needs to set up a two-way communication channel to implement the uplink of FL UEs and the downlink of non-FL UEs.
Second, efficient resource allocation approaches are required at all the steps to control the inter-user interference among FL UEs and non-FL UEs to satisfy their  different service requirements. 

There are two types of communication schemes that are possible to serve the two-way communication between the central server and UEs, namely half-duplex (HD) and full-duplex (FD). Each of these communication schemes has its own advantages and disadvantages \cite{Shende13HDFD}. The main draw back of the FD scheme is the self-interference (SI) between transmit and receive antennas of the BS can cause significant performance degradation, which does not appear in the HD communication. However, for small or moderate SI, the FD communication can approximately double the spectral efficiency compared to the half duplex (HD) scheme \cite{Sabharwal14JSAC}. Both HD and FD schemes are popular in the literature of massive multiple-input
multiple-output (MIMO) networks \cite{Ngo17mMIMO,Sharma18TCOM,Wang17TVT}. However, they cannot be straightforwardly applied to the massive MIMO systems that serve both FL and non-FL UEs.

In this paper, we follow the communication-oriented approach and propose a novel network design for jointly serving FL and downlink non-FL UEs\footnote{The network design for uplink non-FL UEs is open for future works.}
at the same time. First, we propose a communication scheme using massive MIMO and let each FL communication round be executed in one large-scale coherence time.\footnote{Large-scale coherence time is a time interval where the large-scale
fading coefficient remains reasonably invariant.} Because of the high array gain, multiplexing gain, and macro-diversity gain, massive MIMO provides a reliable operation of each FL communication round as well as the whole FL process \cite{vu20TWC}. Here, in the first step of each
FL communication round, both groups are jointly served in the downlink by the central server and in the third step, either of the HD and FD schemes is considered to serve the uplink transmission of FL UEs and the downlink transmission of non-FL UEs. Next, we formulate an optimization problem that optimally allocates power and computation resources to maximize the fairness of effective data rates for non-FL UEs, while ensuring a quality-of-service time of each FL iteration for FL UEs. A successive convex approximation algorithm is then proposed to solve the formulated problem. 
In particular, our contributions are as follows: 
\begin{itemize}
\item We propose HD and FD communication schemes to jointly serve both FL and non-FL UEs in a massive MIMO network, which has not  been studied previously. In the proposed HD scheme, the total system bandwidth is divided equally between the FL and non-FL groups in the uplink of each FL iteration such that both groups are served at the same time in different bandwidths. In the FD communication scheme, both FL UEs and non-FL UEs transmit and receive data in the same time and bandwidth resource under the presence of SI.
\item We propose a new performance measure, called the ``\emph{effective} data'', which is defined as the  amount of data  received by the non-FL UEs, per unit latency time taken by FL UEs.   Then, we formulate an optimization problem to maximize the minimum effective data subject to a QoS constraint on the execution time for FL UEs. Due to the nonconvexity of the formulated problem, we propose a successive convex approximation (SCA) algorithm to find a stationary solution. 
\item We provide an extensive set of simulation results to compare the proposed HD-based and FD-based schemes with two baseline schemes: The first baseline scheme makes use of the frequency division multiple access (FDMA) approach to serve each user independently in an allocated bandwidth, while the second baseline scheme considers an equal power allocation (EPA) approach to find the power control. It is observed that the proposed  HD and FD schemes provide significantly better solution than two considered baseline schemes. Numerical results also show that the FD scheme is a better choice than the HD scheme when the size of the model updates is large and/or when the SI is small or moderate. 
\end{itemize}

\emph{Notation}s: Bold lower and upper case letters represent vectors and matrices, respectively. The notations $\mathbb{R}$ and $\mathbb{C}$ represent the space of real and complex numbers, respectively. $\Vert\cdot\Vert$ represents the Euclidean norm; $|\cdot|$ is the absolute value of the argument. $\mathcal{CN}(0,a)$ denotes a complex Gaussian random variable with zero mean and variance $a$.  $\mathbf{X}^{T}$ and $\mathbf{X}^{H}$ stand for the transpose and Hermitian of $\mathbf{X}$, respectively. The operators $\EEE\{\cdot\}$ and $\var\{\cdot\}$ represent expectation and variance of the argument, respectively. 

\section{System Model and Proposed Transmission Scheme}

\label{sec:SystModel}

\subsection{System Model}

We consider a massive MIMO system where a 
BS serves simultaneously non-FL UEs and FL UEs.  We assume that the non-FL UEs are only those receiving data in the
 downlink transmission.
 Let $\LL\triangleq\{1,\dots,L\}$, and $\K\triangleq\{1,\dots,K\}$
be the sets of FL UEs and non-FL UEs, respectively. 
All  FL and non-FL UEs are equipped with a single antenna, while
the BS has $M$ transmit antennas and $M$ receive antennas. 

To serve FL UEs, the BS acts as a central server. There are four main steps in each iteration of a standard FL framework, i.e., global update downlink transmission, local update computation at the UEs, local update uplink transmission, and global update computation at the BS   \cite{ma15OMS,tran19INFOCOM,mcmahan17AISTATS}. To serve non-FL UEs,  
as mentioned above, the BS constantly transmits downlink data to the non-FL UEs at the same time when all four steps of each FL iteration are executed. Thus, the transmission protocol of our considered system can be summarized as the following four steps in each FL
iteration: 
\begin{enumerate}[label=(S\arabic*)]
\item The BS sends a global update through the downlink channel to
FL UEs. At the same time, non-FL UEs also receive downlink data from the BS. 
\item The FL UEs update their local training model based on the
global update and solve their local learning problems to obtain their
local updates. During this time duration, non-FL UEs continue receiving downlink data from BS.
\item The locally computed updates are sent by FL UEs to the BS
in the uplink channel while the downlink data is still being sent from the BS to non-FL UEs.
\item The BS computes the global update by aggregating the received local updates. 
\end{enumerate}

In Step (S3), we need to serve both FL and non-FL UEs. In this regard, there are two types of possible communication schemes: HD and FD. In the HD scheme, the FL and non-FL groups are served in different frequency bands, while in the FD scheme, both groups are served in the same time and frequency resource.
During Step (S4), the BS computes its global update after receiving all the local update, the delay of computing the global update is negligible since the computational capability of the central server is much more powerful than those of the UEs. Therefore, the downlink amount of data received by the non-FL UEs during the fourth step is not considered in the rest of the paper.

\subsection{Proposed Transmission Schemes}
\begin{figure}[t!]
\centering \includegraphics[width=0.7\textwidth]{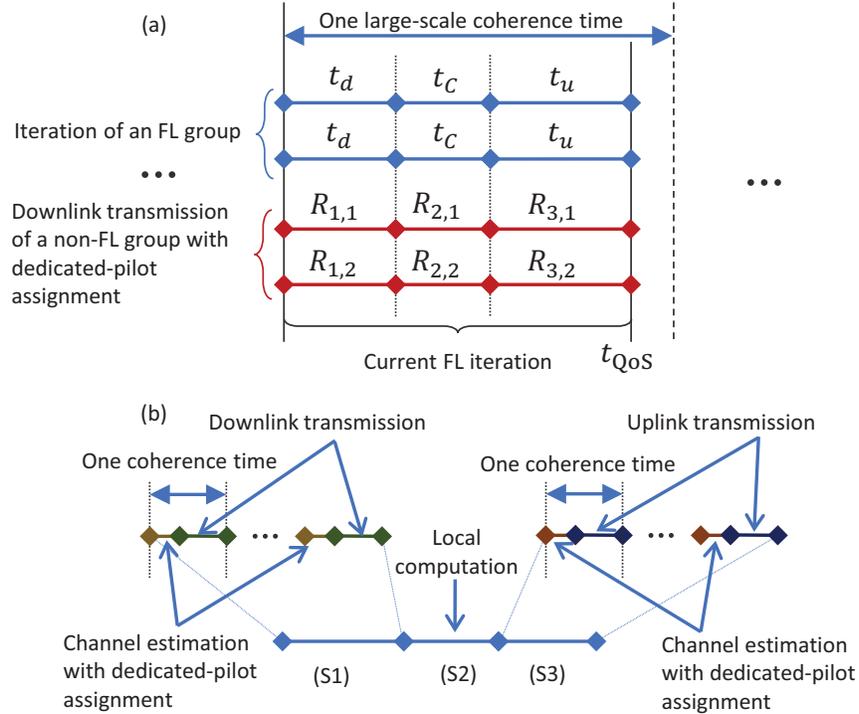}
\vspace{-5mm}
 \caption{(a): Illustration of FL iterations over the considered massive MIMO network
with two groups of FL and non-FL UEs, and two UEs in each group. (b):
Detailed operation of one FL iteration of the FL group.}
\label{fig:time} 
\end{figure}

We propose to use a scheme in \cite{vu20TWC} to support FL iterations as in Fig.~\ref{fig:time}(a). We assume that each FL iteration is executed within a large-scale coherence time. All the FL UEs start each step of
their FL iterations at the same time, and wait for others to finish
their steps before starting a new step. The global and local updates in Steps
(S1) and (S3) are transmitted in multiple (small-scale)
coherence interval, as shown in Fig.~\ref{fig:time}(b).
Each small-scale coherence interval in Step (S1) or (S3) includes two phases: channel estimation and downlink or uplink transmission. In the following, we will provide details of our proposed transmission protocol for both HD and HD modes at the BS in Step (S3).

\subsubsection{Step (S1)}

In this step, the BS wants to send the global updates to all FL UEs
via a downlink transmission while simultaneously sending the payload data to all the
non-FL UEs. 

\textbf{Channel estimation}: The BS estimates the channels by using
uplink pilots received from all the UEs with a time-division-duplexing (TDD) protocol and exploiting
channel reciprocity. Let $\sqrt{\rho_{p}}\boldsymbol{\varphi}_{\ell}\in\mathbb{C}^{\tau_{d,p}\times 1}$, where $\Vert\boldsymbol{\varphi}_{\ell}\Vert^{2}=1$,
be the dedicated pilot symbols assigned to the $\ell$-th FL UE,
and $\sqrt{\rho_{p}}\bar{\boldsymbol{\varphi}}_{k}\in\mathbb{C}^{\tau_{1,p}\times1}$, where $\Vert\bar{\boldsymbol{\varphi}}_{k}\Vert^{2}=1$, 
be the pilot sequence assigned to the $k$-th non-FL UE, where
$\rho_{p}$ is the normalized signal to noise ratio (SNR) of each
pilot vector. In addition, $\tau_{d,p}$ and $\tau_{1,p}$ are the corresponding pilot lengths.
We assume $\tau_{d,p},\tau_{1,p}\geq L+K$, and the pilots of non-FL UEs and FL UEs are pairwisely orthogonal
i.e. $\boldsymbol{\varphi}_{\ell}^{H}\bar{\boldsymbol{\varphi}}_{k}=0,\forall\ell,\forall k$,
$\boldsymbol{\varphi}_{\ell}^{H}\boldsymbol{\varphi}_{\ell^{\prime}}=0,\forall\ell^{\prime}\neq\ell$
and $\bar{\boldsymbol{\varphi}}_{k}^{H}\bar{\boldsymbol{\varphi}}_{k^{\prime}}=0,\forall k^{\prime}\neq k$.

Let $\G_{d}=[\g_{d,1},\dots,\g_{d,L}]\in\mathbb{C}^{M\times L}$ and
$\HH_{1}=[\Hh_{1,1},\dots,\Hh_{1,K}]\in\mathbb{C}^{M\times K}$ be
the channel matrices from the BS to the FL and non-FL groups in
Step (S1), respectively. Here, $\g_{d,\ell}$ represents the channel
vector from the BS to the $\ell$-th FL UE, while $\Hh_{1,k}$ is
the channel vector between the BS and non-FL UE $k$ in Step (S1). We assume Rayleigh fading, i.e., $\g_{d,\ell} \sim\mathcal{CN}(\mathbf{0},\beta_{\ell}\mathbf{I}_{M})$ and $\Hh_{1,k} \sim\mathcal{CN}(\mathbf{0},\bar{\beta}_{k}\mathbf{I}_{M})$, where $\beta_{\ell}$ and $\bar{\beta}_{k}$ represent large-scale fading.
The minimum mean square error (MMSE) estimate  of
$\g_{d,\ell}$  can be written as $\check{\g}_{d,\ell}=\sigma_{d,\ell}\z_{d,\ell}$,
where $\z_{d,\ell}\sim\mathcal{CN}(\mathbf{0},\mathbf{I}_{M})$, and   $\sigma_{d,\ell}^{2}=\frac{\rho_{p}\tau_{d,p}\beta_{\ell}^{2}}{\rho_{p}\tau_{d,p}\beta_{\ell}+1}$. Similarly,
the MMSE estimate  of $\Hh_{1,k}$  can be
written as $\check{\Hh}_{1,k}=\sigma_{1,k}\z_{1,k}$, where $\z_{1,k}\sim\CN(\mathbf{0},\mathbf{I}_{M})$, and $\sigma_{1,k}^{2}=\frac{\rho_{p}\tau_{1,p}\bar{\beta}_{k}^{2}}{\rho_{p}\tau_{1,p}\bar{\beta}_{k}+1}$.
Let $\check{\G}_{d}=[\check{\g}_{d,1},\dots,\check{\g}_{d,L}]$, $\check{\HH}_{1}=[\check{\Hh}_{1,1},\dots,\check{\Hh}_{1,K}]$,
$\Z_{d}=[\z_{d,1},\dots,\z_{d,L}]$, $\Z_{1}=[\z_{1,1},\dots,\z_{1,K}]$,
$\SIgma_{d}\triangleq[\sigma_{d,1},\dots,\sigma_{d,L}]^{T}$, and $\SIgma_{1}\triangleq[\sigma_{1,1},\dots,\sigma_{1,K}]^{T}$.
Denote by $\E_{d}=[\Epsilon_{d,1},\dots,\Epsilon_{d,L}]$ and $\E_{1}=[\Epsilon_{1,1},\dots,\Epsilon_{1,K}]$
be the channel estimate error matrices of $\G_{d}$ and $\HH_{1}$, i.e.,
$\Epsilon_{d,\ell}=\check{\g}_{d,\ell}-\g_{d,\ell}$ and $\Epsilon_{1,k}=\check{\Hh}_{1,k}-\Hh_{1,k}$.
From the property
of MMSE estimation, we have that $\Epsilon_{d,\ell}$, $\check{\g}_{d,\ell}$, $\Epsilon_{1,k}$,
and $\check{\Hh}_{1,k}$ are independent, and hence, $\Epsilon_{d,\ell}\sim\CN(\mathbf{0},(\beta_{\ell}-\sigma_{d,\ell}^{2})\mathbf{I}_{M})$,
$\Epsilon_{1,k}\sim\CN(\mathbf{0},(\bar{\beta}_{k}-\sigma_{1,k}^{2})\mathbf{I}_{M})$.

\textbf{Downlink transmission for both FL and non-FL UEs}: The BS
encodes downlink data desired for non-FL UE $k$  into the symbol
$s_{1,k}\sim\CN(0,1), \forall k\in\K$, and the global training update intended for the FL UE $\ell$ into symbol
$s_{d,\ell}\sim\CN(0,1), \forall\ell\in\mathcal{L}$. Note that the global update is the same for all FL UEs but we use different coding schemes for different UEs. The zero-forcing (ZF) precoding
scheme is then applied to precode the symbols for FL and non-FL groups. Let $\s_{d}\triangleq[s_{d,1},\dots,s_{d,L}]^{T}$, $\s_{1}\triangleq[s_{1,1},\dots,s_{1,K}]^{T}$. With ZF, $M\geq L+K$ is required, and the signal transmitted  at the BS in Step (S1) is  given by $$\x_{1}=\sqrt{\rho_{d}}\U_{d}\D_{\ETA_{d}}^{1/2}\s_{d}+\sqrt{\rho_{d}}\U_{1}\D_{\ZETA_{1}}^{1/2}\s_{1},$$
where $[\U_{d} ~ \U_{1}]=\sqrt{(M-L-K)}\Z(\Z^{H}\Z)^{-1}$
\cite[(3.49)]{ngo16}, with $\Z=[\Z_{d},\Z_{1}]$. In addition, $\D_{\ETA_{d}}$ and $\D_{\ZETA_{1}}$ are diagonal matrices with the elements of $\ETA_{d}$ and $\ZETA_{1}$ on their
diagonal, respectively, where the $\ell$-th element of $\ETA_{d}$ denoted by $\eta_{d,\ell}$ and the $k$-th element of $\ZETA_{1}$ denoted by $\zeta_{1,k}$ are the power control coefficients
associated with the $\ell$-th FL UE and $k$-th non-FL UE, respectively. The transmitted power at the BS is required to meet the average normalized
power constraint, i.e., $\EEE\{|\mathbf{x}_{1}|^{2}\}\leq\rho_{d}$,
which can be expressed as: 
\begin{equation}
\sum\nolimits _{\ell\in\mathcal{L}}\eta_{d,\ell}+\sum\nolimits _{k\in\K}\zeta_{1,k}\leq1.\label{powerdupperbound}
\end{equation}
The received signal vector collected from all FL UEs is given by 
\begin{align}
\y_{d} & =\G_{d}^{H}\x_{1}+\n_{d}=\sqrt{\rho_{d}}{\G}_{d}^{H}\U_{d}\D_{\ETA_{d}}^{1/2}\s_{d}+\sqrt{\rho_{d}}\G_{d}^{H}\U_{1}\D_{\ZETA_{1}}^{1/2}\s_{1}+\n_{d},
\end{align}
where $\n_{d}\sim\mathcal{CN}(\mathbf{0},\mathbf{I}_{L})$ is the
additive noise. Since $\check{\G}_{d}^{H}\U_{d}=\sqrt{M-L-K} \D_{\pmb{\sigma}_{d}}$ and $\check{\G}_{d}^{H}\U_{1}=\mathbf{0}$, the $\ell$-th FL UE receives 
\begin{align}
y_{d,\ell} & =\sqrt{\rho_{d}\eta_{\ell}(M-L-K)}\sigma_{d,\ell}s_{d,\ell}+n_{d,\ell}-\sqrt{\rho_{d}}\boldsymbol{\epsilon}_{d,\ell}^{H}\U_{d}\D_{\ETA_{d}}^{1/2}\s_{d}-\sqrt{\rho_{d}}\boldsymbol{\epsilon}_{d,\ell}^{H}\U_{1}\D_{\ZETA_{1}}^{1/2}\s_{1}.
\end{align}
Following \cite[Sec. 3.3.2]{ngo16}, the effective SINR at the
$\ell$-th FL UE is given by 
\begin{align}
 & \text{SINR}_{d,\ell}(\ETA_{d},\ZETA_{1})=\frac{\rho_{d}\eta_{d,\ell}(M-L-K)\sigma_{d,\ell}^{2}}{1+\rho_{d}\var\{\boldsymbol{\epsilon}_{d,\ell}^{H}\U_{d}\D_{\ETA_{d}}^{1/2}\s_{d}+\boldsymbol{\epsilon}_{d,\ell}^{H}\U_{1}\D_{\ZETA_{1}}^{1/2}\s_{1}\}}.\label{eq:SINR_FL_S1}
\end{align}
Since $\Epsilon_{d,\ell}$ is independent of $\U_{d}\D_{\ETA_{d}}^{1/2}\s_{d}$
and  $\U_{1}\D_{\ZETA_{1}}^{1/2}\s_{1}$,
we get the closed-form expression for $\text{SINR}_{d,\ell}$ as 
\begin{align}
 & \text{SINR}_{d,\ell}(\ETA_{d},\ZETA_{1})=\frac{\rho_{d}\eta_{d,\ell}(M-L-K)\sigma_{d,\ell}^{2}}{1+\rho_{d}(\beta_{\ell}-\sigma_{d,\ell}^{2})\sum_{i\in\LL}\eta_{d,i}+\rho_{d}(\beta_{\ell}-\sigma_{d,\ell}^{2})\sum_{k\in\K}\zeta_{1,k}}.\label{eq:SINR_FL_S1.1}
\end{align}
Similarly, the received signal vector combined from all the non-FL UEs in Step
(S1) is given by 
\begin{align}
\y_{1} & =\HH_{1}^{H}\x_{1}+\n_{1}
=\sqrt{\rho_{d}}{\HH}_{1}^{H}\U_{1}\D_{\ZETA_{1}}^{1/2}\s_{1}+\sqrt{\rho_{d}}\HH_{1}^{H}\U_{d}\D_{\ETA_{d}}^{1/2}\s_{d}+\n_{1},
\end{align}
where $\n_{1}\sim\mathcal{CN}(\mathbf{0},\mathbf{I}_{K})$ is the
additive noise. Using the fact that $\check{\HH}_{1}^{H}\U_{1}=\sqrt{M-L-K}\D_{\pmb{\sigma}_{1}}$ and $\check{\HH}_{1}^{H}\U_{d}=\mathbf{0}$, the effective SINR of the $k$-th non-FL UE is given by 
\begin{align}
\text{SINR}_{1,k}(\ETA_{d},\ZETA_{1}) & =\frac{\rho_{d}\zeta_{1,k}(M-L-K)\sigma_{1,k}^{2}}{1+\rho_{d}\var\{\Epsilon_{1,k}^{H}\U_{1}\D_{\ZETA_{1}}^{1/2}\s_{1}+\Epsilon_{1,k}^{H}\U_{d}\D_{\ETA_{d}}^{1/2}\s_{d}\}}\nonumber \\
 & =\frac{\rho_{d}\zeta_{1,k}(M-L-K)\sigma_{1,k}^{2}}{1+\rho_{d}(\bar{\beta}_{k}-\sigma_{1,k}^{2})\sum_{i\in\K}\zeta_{1,i}+\rho_{d}(\bar{\beta}_{k}-\sigma_{1,k}^{2})\sum_{\ell\in\LL}\eta_{d,\ell}}.\label{eq:SINR_NFL_S1}
\end{align}

Since all the FL UEs starts and end a step together, the
achievable rate (bps) of each FL UE is the minimum
achievable rate of the FL group, i.e., 
\begin{align}
\!\!\!R_{d}(\ETA_{d},\ZETA_{1})\!= & \min_{\ell\in\LL}R_{d,\ell}(\ETA_{d},\ZETA_{1})
\triangleq  \min_{\ell\in\LL}\frac{\tau_{c}\!-\!\tau_{d,p}}{\tau_{c}}B\log_{2}\!\big(1\!+\!\text{SINR}_{d,\ell}(\ETA_{d},\ZETA_{1})\big),
\end{align}
where $B$ is the bandwidth and $\tau_{c}$ is the coherence interval. The achievable rate of non-FL UE $k$
is given by 
\begin{equation}
R_{1,k}(\ETA_{d},\ZETA_{1})\!=\!\frac{\tau_{c}\!-\!\tau_{1,p}}{\tau_{c}}B\log_{2}\big(1\!+\!\text{SINR}_{1,k}(\ETA_{d},\ZETA_{1})\big).\label{R1k}
\end{equation}

\textbf{Downlink delay of the FL group}: Let $S_{d}$ (bits) be the
data size of the global training update of the FL group. The transmission
time from the BS to FL UE $\ell\in\LL$ is given by 
\begin{equation}
t_{d}(\ETA_{d},\ZETA_{1})=\frac{S_{d}}{R_{d}(\ETA_{d},\ZETA_{1})},\forall\ell.\label{eq:TimeFLStep1}
\end{equation}

\textbf{Amount of downlink data received at the non-FL UEs}:

The amount of downlink data received at non-FL UE $k\in\K$ is 
\begin{align}
D_{1,k}(\ETA_{d},\ZETA_{1})=R_{1,k}(\ETA_{d},\ZETA_{1})t_{d}(\ETA_{d},\ZETA_{1}).\label{eq:DataNonFLStep1}
\end{align}

\subsubsection{Step (S2)}

After receiving the global update, each FL UE $\ell$ computes its
local training update on its local dataset, while each non-FL UE $k$
keeps receiving data from the BS.

\textbf{Local computation}: Each
FL UE executes $N_{c}$ local computing rounds over its data set to
compute its local update. Let $c_{\ell}$ (cycles/sample) be the number
of processing cycles for a UE $\ell$ to process one data sample \cite{tran19INFOCOM}.
Denote by $D_{\ell}$ (samples) and $f_{\ell}$ (cycles/s) the size
of the local data set and the processing frequency of  UE $\ell$,
respectively. To provide a certain synchronization in this step, we
choose $f_{\ell}=\frac{D_{\ell}c_{\ell}f}{D_{\max}c_{\max}}$, where $D_{\max}=\max_{\ell\in\LL}D_{\ell}$,
$c_{\max}=\max_{\ell\in\LL}c_{\ell}$, and $f$ is a frequency control
coefficient. The computation time at all the FL UEs of the FL group
is the same $t_{C}(f)$, which is given by \cite{vu20TWC,tran19INFOCOM}
\begin{equation}
\!\!t_{C}(f)\!=\!t_{C,\ell}(f)\!=\!\frac{N_{c}D_{\ell}c_{\ell}}{f_{\ell}}\!=\!\frac{N_{c}D_{\max}c_{\max}}{f},\forall\ell\in\LL.\label{eq:TimeFLStep2}
\end{equation}

\textbf{Channel estimation for non-FL UEs channel}: In Step (S2),
the channel estimation is performed similarly to Step (S1) for the non-FL
UEs. The MMSE estimate 
of $\Hh_{2,k}$ (the channel between the BS and non-FL UE $k$ in Step (S2) can be written as $\check{\Hh}_{2,k}=\sigma_{2,k}\z_{2,k}$,
where $\z_{2,k}\sim\mathcal{CN}(\mathbf{0},\mathbf{I}_{M})$ and $\sigma_{2,k}^{2}=\frac{\rho_{p}\tau_{2,p}\bar{\beta}_{k}^{2}}{\rho_{p}\tau_{2,p}\bar{\beta}_{k}+1}$.
where $\tau_{2,p}\geq K$  is the length of pilot sequence
 in Step (S2).

\textbf{Amount of downlink data received at the non-FL group}: 
Similarly to Step (S1), ZF is used at the BS to transmit signals to $K$ non-FL UEs. Let $\ZETA_{2}\triangleq[\zeta_{2,1},\dots,\zeta_{2,K}]^{T}$ be the
power control coefficients for non-FL UEs. The transmitted
power at the BS is required to meet the average normalized power constraint
which can be expressed as: 
\begin{equation}
\sum\nolimits _{k\in\K}\zeta_{2,k}\leq1.\label{powerdupperbound-1}
\end{equation}
The achievable downlink rate (bps) of non-FL UE $k,\forall k\in\K,$
is given by \cite[(3.49)]{ngo16}
\begin{align}
R_{2,k}(\ZETA_{2}) & =\frac{\tau_{c}-\tau_{2,p}}{\tau_{c}}B\log_{2}\big(1+\text{SINR}_{2,k}(\ZETA_{2})\big),
\end{align}
where $\text{SINR}_{2,k}(\ZETA_{2})$ is the effective SINR  given as 
\begin{align}
\text{SINR}_{2,k}(\ZETA_{2})= & \frac{\rho_{d}\zeta_{2,k}(M-K)\sigma_{2,k}^{2}}{1+\rho_{d}(\bar{\beta}_{k}-\sigma_{2,k}^{2})\sum_{i\in\K}\zeta_{2,i}}.\label{eq:SINR_NFL_S2}
\end{align}
The above equation is similar to \eqref{eq:SINR_NFL_S1} except that
there is no interference induced by FL UEs in Step (S2). 
Thus, the total amount of downlink data received at  non-FL
UE $k$ is 
\begin{equation}
D_{2,k}(\ZETA_{2},f)=R_{2,k}(\ZETA_{2})t_{C}(f).\label{eq:DataNonFLStep2}
\end{equation}

\subsubsection{Step (S3) using HD\label{subsec:Step-(S3)}}
In Step (S3), FL UEs' local updates are transmitted to the BS while
data is kept being sent from the BS to the non-FL UEs. To serve
both the FL and non-FL UEs, two types of duplex communication schemes are possible: HD and FD operations. Using HD in Step (S3),
the system bandwidth is equally divided between the FL and non-FL
groups.

\textbf{Channel estimation}: In Step (S3), the channels between the
BS and FL UEs are estimated using the MMSE estimation technique similarly to Steps (S1)
and (S2). The channel $\g_{u,\ell}$ between the BS and  FL UEs
$\ell$ in Step (S3) has an estimate as $\check{\g}_{u,\ell}=\sigma_{u,\ell}\z_{u,\ell}$,
where $\z_{u,\ell}\sim\mathcal{CN}(\mathbf{0},\mathbf{I}_{M})$  and $\sigma_{u,\ell}^{2}=\frac{\rho_{p}\tau_{u,p}\beta_{\ell}^{2}}{\rho_{p}\tau_{u,p}\beta_{\ell}+1}$, where $\tau_{u,p}\geq L+K$ is the pilot length.
The channel $\Hh_{3,k}$ between the BS and the $k$-th non-FL UEs
in Step (S3) has an estimate $\check{\Hh}_{3,k}=\sigma_{3,k}\z_{3,k}$,
where $\z_{3,k}\sim\mathcal{CN}(\mathbf{0},\mathbf{I}_{M})$ and $\sigma_{3,k}^{2}=\frac{\rho_{p}\tau_{3,p}\bar{\beta}_{k}^{2}}{\rho_{p}\tau_{3,p}\bar{\beta}_{k}+1}$.
Here, $\tau_{3,p}\geq L+K$ is the length of pilot sequence
 in Step (S3). Let $\Z_{u}\triangleq[\z_{u,1},\dots,\z_{u,L}]$,
$\Z_{3}\triangleq[\z_{3,1},\dots,\z_{3,K}]$, $\G_{u}\triangleq[\g_{u,1},\dots,\g_{u,L}]$,
$\check{\G}_{u}=[\check{\g}_{u,1},\dots,\check{\g}_{u,1}]$, $\HH_{3}\triangleq[\Hh_{3,1},\dots,\Hh_{3,K}]$,
and $\check{\HH}_{3}=[\check{\Hh}_{3,1},\dots,\check{\Hh}_{3,K}]$.
Denote by $\E_{u}=[\Epsilon_{u,1},\dots,\Epsilon_{u,L}]$ and $\E_{3}=[\Epsilon_{3,1},\dots,\Epsilon_{3,K}]$
be the channel estimate error matrices of $\G_{u}$ and $\HH_{3}$, i.e.,
$\E_{d}=\check{\G}_{d}-\G_{d}$ and $\E_{3}=\check{\HH}_{3}-\HH_{3}$.
Here, $\Epsilon_{u,\ell}\sim\CN(\mathbf{0},(\beta_{\ell}-\sigma_{u,\ell}^{2})\mathbf{I}_M)$,
$\check{\g}_{u,\ell}$, $\Epsilon_{3,k}\sim\CN(\mathbf{0},(\bar{\beta}_{k}-\sigma_{3,k}^{2})\mathbf{I}_M)$,
and $\check{\Hh}_{3,k}$ are independent.

\textbf{Uplink transmission of FL UEs}: After computing the local
update, all FL UEs transmit their local updates to the BS. The signal transmitted from FL UE $\ell$ is $$x_{u,\ell}=\sqrt{\rho_{u}\eta_{u,\ell}}s_{u,\ell},$$
 where $s_{u,\ell}\sim\CN(0,1)$ is the data symbol, $\eta_{u,\ell}$ is the power control coefficient chosen to satisfy the average transmit power constraint, i.e., $\EEE\left\{ |x_{u,\ell}|^{2}\right\} \leq\rho_{u}$,
which can be expressed as 
\begin{equation}
\eta_{u,\ell}\leq1,\forall\ell\in\LL.\label{poweruupperbound}
\end{equation}
The received signal vector at the BS is then given as 
\begin{align}
\y_{u}^{\text{HD}} & =\sqrt{\rho_{u}}\G_{u}\D_{\ETA_{u}}^{1/2}\s_{u}+\n_{u},
\end{align}
where $\ETA_{u}=[\eta_{u,1},\dots,\eta_{u,L}]^{T}$ and $\n_{u}\sim\mathcal{CN}(\mathbf{0},\mathbf{I}_{M})$
is the additive noise vector. 

After receiving signals from all the UEs, the BS applies a ZF decoding
scheme for detecting the FL UEs' symbols. With ZF, signal used for detecting $s_{u,\ell}$ is given by 
\begin{align}
y_{u,\ell}^{\text{HD}}= & \sqrt{\rho_{u}}\uu_{u,\ell}^{H}\G_{u}\D_{\ETA_{u}}^{1/2}\s_{u}+\uu_{u,\ell}^{H}\n_{u}\nonumber \\
= & \sqrt{\rho_{u}}\uu_{u,\ell}^{H}\check{\G}_{u}\D_{\ETA_{u}}^{1/2}\s_{u}-\sqrt{\rho_{u}}\uu_{u,\ell}^{H}\E_{u}\D_{{\boldsymbol{\eta}}_{u}}^{1/2}{\s}_{u}+\uu_{u,\ell}^{H}\n_{u}\nonumber \\
= & \sqrt{\rho_{u}\eta_{u,\ell}(M-L)}\sigma_{u,\ell}s_{u,\ell}-\sqrt{\rho_{u}}\uu_{u,\ell}^{H}\E_{u}\D_{\ETA_{u}}^{1/2}\s_{u}+\uu_{u,\ell}^{H}\n_{u},
\end{align}
where $\uu_{u,\ell}=\sqrt{(M-L)}\Z_{u}(\Z_{u}^{H}\Z_{u})^{-1}\mathbf{e}_{\ell,L}$
is the zero-forcing decoding vector. For synchronization, we choose
the rates of FL UEs to be the same as the minimum achievable rates
in the FL group, i.e., 
\begin{align}
R_{u}^{\text{HD}}(\ETA_{u}) & =\min_{\ell\in\LL}R_{u,\ell}^{\text{HD}}(\ETA_{u})
 \triangleq\min_{\ell\in\LL}\frac{\tau_{c}-\tau_{u,p}}{\tau_{c}}\frac{B}{2}\log_{2}\big(1+\text{SINR}_{u,\ell}^{\text{HD}}(\ETA_{u})\big),\label{eq:uplinkRate}
\end{align}
where $1/2$ appears in the pre-log factor of the rate comes from the fact that the system bandwidth is equally divided between the FL and non-FL groups, and
\begin{align}
\text{SINR}_{u,\ell}^{\text{HD}}(\ETA_{u})=\frac{\rho_{u}\eta_{u,\ell}(M-L)\sigma_{u,\ell}^{2}}{1\!+\!\rho_{d}\var\{\uu_{u,\ell}^{H}\E_{u}\D_{\ETA_{u}}^{1/2}\s_{u}\!\}}.\label{eq:SINR_FL_S3}
\end{align}
The above equation is then computed as 
\begin{align}
\text{SINR}_{u,\ell}^{\text{HD}}(\ETA_{u})=\frac{\rho_{u}\eta_{u,\ell}(M-L)\sigma_{u,\ell}^{2}}{1+\rho_{u}\sum_{i\in\LL}(\beta_{u,i}-\sigma_{u,i}^2)\eta_{u,i}}.\label{eq:SINR_FL_S3-2}
\end{align}

\textbf{Downlink transmission for Non-FL UEs}: Denote by $\s_{3} = [s_{3,1} ~ \ldots ~ s_{3,K}]^T$ the vector of $K$ symbols intended for $K$ non-FL UEs, and $\U_{3}=\sqrt{(M-K)}\Z_{3}(\Z_{3}^{H}\Z_{3})^{-1}$
 the ZF precoding matrix. Then, the transmitted signal from the BS to the non-FL UEs is given as $$\x_{3}=\sqrt{\rho_{d}}\U_{3}\D_{\ZETA_{3}}^{1/2}\s_{3},$$
where $\ZETA_{3}\triangleq[\zeta_{3,1},\dots,\zeta_{3,K}]^{T}$, and $\zeta_{3,k}$
the power control coefficient allocated for non-FL UE $k$
chosen to meet the average normalized
power constraint at the BS, i.e., $\EEE\{|\x_{3}|^{2}\}\leq\rho_{d}$, which
can be expressed as: 
\begin{equation}
\sum_{k\in\K}\zeta_{3,k}\leq1.\label{powerdupperbound-2}
\end{equation}
For the $k$-th non-FL UE, the received signal can be written as 
\begin{align}
y_{3,k}^{\text{HD}}= & \sqrt{\rho_{d}}\check{\Hh}_{3,k}^{H}\U_{3}\D_{\ZETA_{3}}^{1/2}\s_{3}-\sqrt{\rho_{d}}\Epsilon_{3,k}^{H}\U_{3}\D_{\ZETA_{3}}^{1/2}\s_{3}+n_{3,k}\nonumber \\
= & \sqrt{\rho_{d}\eta_{3,k}(M-K)}\sigma_{3,k}s_{3,k}-\sqrt{\rho_{d}}\Epsilon_{3,k}^{H}\U_{3}\D_{\ZETA_{3}}^{1/2}\s_{3}+n_{3,k}.
\end{align}
In the above equation, the term $\Epsilon_{3,k}$ is independent of $\U_{3}\D_{\ZETA_{u}}^{1/2}\s_{3}$.
 Thus, under HD in Step (S3), the effective
SINR for the downlink payload at non-FL UE $k$ is
\begin{align}
\text{SINR}_{3,k}^{\text{HD}}(\ZETA_{3}) & =\frac{\rho_{d}\zeta_{3,k}(M-K)\sigma_{3,k}^{2}}{1+\rho_{d}\var\{\Epsilon_{3,k}^{H}\U_{3}\D_{\ZETA_{3}}^{1/2}\s_{3}\}}
 =\frac{\rho_{d}\zeta_{3,k}(M-K)\sigma_{3,k}^{2}}{1+\rho_{d}(\bar{\beta}_{k}-\sigma_{3,k}^2)\sum_{i\in\K}\zeta_{3,i}},\label{eq:SINR_NFL_S3}
\end{align}
and the achievable downlink rate for non-FL UE $k,\forall k\in\K,$
is 
\begin{align}
R_{3,k}^{\text{HD}}(\ZETA_{3}) & \!=\!\frac{\tau_{c}\!-\!\tau_{3,p}}{\tau_{c}}\frac{B}{2}\log_{2}\big(1\!+\!\text{SINR}_{3,k}(\ZETA_{3})\big).
\end{align}

\textbf{Uplink delay}: Denote by $S_{u}$ (bits) the data size of
the local training update of the FL group. The transmission time from
each FL UE to the BS is the same and given by 
\begin{equation}
t_{u}^{\text{HD}}(\ETA_{u})=\frac{S_{u}}{R_{u}^{\text{HD}}(\ETA_{u})}.\label{eq:TimeFLStep3}
\end{equation}

\textbf{Amount of downlink data received at the non-FL group}: The
amount of downlink data received at the non-FL UE $k,\forall k\in\K$,
in Step (S3) using HD is 
\begin{equation}
D_{3,k}^{\text{HD}}(\ETA_{u},\ZETA_{3})=R_{3,k}^{\text{HD}}(\ZETA_{3})t_{u}^{\text{HD}}(\ETA_{u}).\label{eq:DataNonFLStep3}
\end{equation}

\subsubsection{Step (S3) using FD}

Step  (S3) involves transmission in both directions. This motivates us to consider the FD communications to serve both
groups of UEs simultaneously. Specifically,  FL UEs send
their local updates to the BS in the uplink channel and at the same
time, non-FL UEs receive the downlink data from the BS.  The proposed FD scheme is detailed in what follows.

\textbf{Uplink transmission of FL UEs}: In the FD communications, channel
coefficients are estimated similarly to what was done in case of the
HD communications. FL UEs transmit the locally computed updates to
the BS in the presence of non-FL UEs which are receiving the downlink
data. Therefore, SI is present between the receiver
and transmit antennas of the BS which is denoted by $\G^{\mathrm{SI}}\in\mathbb{C}^{M\times M}$.
The elements of matrix $\G_{\mathrm{SI}}$ are modeled as i.i.d random
variables and are given by $\sigma_{\text{SI}}^{2}=\beta_{\text{SI}}\sigma_{\text{SI},0}^{2},$
where $\beta_{\text{SI}}$ represents  the pass loss
from a transmit antenna to a receive antenna of the BS due to their physical antenna seperation and $\sigma_{\text{SI},0}^{2}$
is the power of the residual interference at each BS antenna after
the SI suppression, respectively. Similar to the HD scheme, the baseband
signal is subjected to the average transmit power constraint \eqref{poweruupperbound}.
The received signal vector at the BS in case of FD communication is
expressed as
\begin{align}
\y_{u}^{\text{FD}} & =\sqrt{\rho_{u}}\G_{u}\D_{\ETA_{u}}^{1/2}\s_{u}+\sqrt{\rho_{d}}\G_{\mathrm{SI}}\U_{3}\D_{\ZETA_{3}}^{1/2}\s_{3}+\n_{u},
\end{align}
where $\n_{u}\sim\mathcal{CN}(\mathbf{0},\mathbf{I}_{M})$ is the
vector of additive noise components. Note that SI
is caused from transmit antennas of the BS to receiving antennas and
thus, the effective noise has an additional SI term caused by the
downlink transmission to non-FL UEs. After receiving signals from
all the UEs, the BS applies a ZF decoding scheme for detecting the
FL UEs' symbols. The detected signal for the $\ell$-th FL UE is given
by 
\begin{align}
y_{u,\ell}^{\text{FD}}= & \sqrt{\rho_{u}}\uu_{u,\ell}^{H}\G_{u}\D_{\ETA_{u}}^{1/2}\s_{u}+\sqrt{\rho_{d}}\uu_{u,\ell}^{H}\G_{\mathrm{SI}}\U_{3}\D_{\ZETA_{3}}^{1/2}\s_{3}+\uu_{u,\ell}^{H}\n_{u}\nonumber \\
= & \sqrt{\rho_{u}}\uu_{u,\ell}^{H}\check{\G}_{u}\D_{\ETA_{u}}^{1/2}\s_{u}-\sqrt{\rho_{u}}\uu_{u,\ell}^{H}\E_{u}\D_{{\boldsymbol{\eta}}_{u}}^{1/2}{\s}_{u}+\sqrt{\rho_{d}}\uu_{u,\ell}^{H}\G_{\mathrm{SI}}\U_{3}\D_{\ZETA_{3}}^{1/2}\s_{3}+\uu_{u,\ell}^{H}\n_{u}\nonumber \\
= & \sqrt{\rho_{u}\eta_{u,\ell}(M-L)}\sigma_{u,\ell}s_{u,\ell}-\sqrt{\rho_{u}}\uu_{u,\ell}^{H}\E_{u}\D_{\ETA_{u}}^{1/2}\s_{u}+\sqrt{\rho_{d}}\uu_{u,\ell}^{H}\G_{\mathrm{SI}}\U_{3}\D_{\ZETA_{3}}^{1/2}\s_{3}+\uu_{u,\ell}^{H}\n_{u}.
\end{align}
The SINR for the $\ell$-th FL UE in case of FD communications is given
as 
\begin{align}
\text{SINR}_{u,\ell}^{\text{FD}}(\ETA_{u},\ZETA_{3}) & =\frac{\rho_{u}\eta_{u,\ell}(M-L)\sigma_{u,\ell}^{2}}{1\!+\!\rho_{d}\var\{\uu_{u,\ell}^{H}\E_{u}\D_{\ETA_{u}}^{1/2}\s_{u}\!+\!\uu_{u,\ell}^{H}\G_{\mathrm{SI}}\U_{3}\D_{\ZETA_{3}}^{1/2}\s_{u}\!\}}.\label{eq:SINR_FL_S3-1}
\end{align}

\begin{prop}
The SINR for the $\ell$-th FL UE in case of FD communications given in \eqref{eq:SINR_FL_S3-1} can be approximated by
\end{prop}
\vspace{-10mm}
\begin{align}
\text{SINR}_{u,\ell}^{\text{FD}}(\ETA_{u},\ZETA_{3})\!\approx\! \widehat{\text{SINR}}_{u,\ell}^{\text{FD}}(\ETA_{u},\ZETA_{3})\!
=\!
\frac{\rho_{u}\eta_{u,\ell}(M-L)\sigma_{u,\ell}^{2}}{1\!+\!\rho_{u}\sum_{i\in\LL}(\beta_{u,i}\!-\!\sigma_{u,i}^2)\eta_{u,i}\!+\!\rho_{d}M\beta_{\text{SI}}\sigma_{\mathrm{SI},0}^{2}\sum_{j\in\K}\zeta_{3,j}}.\label{eq:SINR_FL_S3-2-2}
\end{align}

\begin{IEEEproof}
Proof of \eqref{eq:SINR_FL_S3-2-2} is provided in Appendix \ref{sec:FDrates}.
\end{IEEEproof}
For synchronization, we again choose the rates of FL UEs to be the
same as the minimum achievable rates in the FL group.
\begin{align}
R_{u}^{\text{FD}}(\ETA_{u},\ZETA_{3}) & =\min_{\ell\in\LL}R_{u,\ell}^{\text{FD}}(\ETA_{u},\ZETA_{3}) 
  \triangleq\min_{\ell\in\LL}\frac{\tau_{c}-\tau_{u,p}}{\tau_{c}}B\log_{2}\big(1+\widehat{\text{SINR}}_{u,\ell}^{\text{FD}}(\ETA_{u},\ZETA_{3})\big).\label{eq:uplinkRate-1}
\end{align}
Above equation is similar to \eqref{eq:uplinkRate} except that FL
UEs make use to the full bandwidth in the FD communication.

\textbf{Downlink transmission for Non-FL UEs}: In FD, non-FL UEs continue
receiving data from the BS in the downlink channel in the presence
of FL UEs which simultaneously send the local updates to the BS in
the uplink channel. Therefore, the received signal at each non-FL
UE contains the inter-group interference (IGI) from the group of FL
UEs. To approximate the SINR in this case, the transmitted power at
the BS is constrained to meet the average normalized power constraint
\eqref{powerdupperbound-2} similar to the HD scheme.

The received signal for the $k$-th non-FL UE can be written as 
\begin{align}
y_{3,k}^{\text{FD}}= & \sqrt{\rho_{d}}\Hh_{3,k}^{H}\U_{3}\D_{\ZETA_{3}}^{1/2}\s_{3}+\sqrt{\rho_{u}}\HH_{\mathrm{IGI}}\D_{\ETA_{u}}^{1/2}\s_{u}+n_{3,k},
\end{align}
where $\HH_{\mathrm{IGI}}\in\mathbb{C}^{L\times K}$ be the inter-group
channel matrix whose elements are modeled as $h_{\mathrm{IGI},k\ell}=\beta_{\mathrm{IGI},k\ell}^{1/2}\bar{h}_{\mathrm{IGI},k\ell}$, where $\beta_{\mathrm{IGI},k\ell}$ is the large-scale fading and $\bar{h}_{\mathrm{IGI},k\ell}\sim\mathcal{CN}(0,1)$ is the small-scale fading of the inter-group channel.
After the channel estimation, the first term in the above equation
can be broken into the estimation term and the error term and thus,
the above equation can be rewritten as
\begin{align}
y_{3,k}^{\text{FD}}= & \sqrt{\rho_{d}}\check{\Hh}_{3,k}^{H}\U_{3}\D_{\ZETA_{3}}^{1/2}\s_{3}-\sqrt{\rho_{d}}\Epsilon_{3,k}^{H}\U_{3}\D_{\ZETA_{3}}^{1/2}\s_{3}+\sqrt{\rho_{u}}\HH_{\mathrm{IGI}}\D_{\ETA_{u}}^{1/2}\s_{u}+n_{3,k}\\
= & \sqrt{\rho_{d}\eta_{3,k}(M-K)}\sigma_{3,k}s_{3,k}-\sqrt{\rho_{d}}\Epsilon_{3,k}^{H}\U_{3}\D_{\ZETA_{3}}^{1/2}\s_{3}+\sqrt{\rho_{u}}\HH_{\mathrm{IGI}}\D_{\ETA_{u}}^{1/2}\s_{u}+n_{3,k}.
\end{align}
The effective SINR in the downlink payload at non-FL UE $k$ is given
as 
\begin{align}
\text{SINR}_{3,k}^{\text{FD}}(\ETA_{u},\ZETA_{3}) & =\frac{\rho_{d}\eta_{3,k}(M-K)\sigma_{3,k}^{2}}{1+\var\{\rho_{d}\Epsilon_{3,k}^{H}\U_{3}\D_{\ZETA_{3}}^{1/2}\s_{3}+\rho_{u}\HH_{\mathrm{IGI},k}^{H}\D_{\ETA_{u}}^{1/2}\s_{u}\}}.\label{eq:SINR_NFL_S3-1}
\end{align}
Note that in the above equation, $\Epsilon_{3,k}$ is independent
of $\U_{3}\D_{\ZETA_{u}}^{1/2}\s_{3}$. Moreover, $\var\{\rho_{u}\HH_{\mathrm{IGI},k}^{H}\D_{\ETA_{u}}^{1/2}\s_{u}\}$
simplifies to $\rho_{u}\sum_{i\in\LL}\eta_{u,i}\beta_{\mathrm{IGI},ki}$.
Thus, the effective SINR can be rewritten as 
\begin{align}
\text{SINR}_{3,k}^{\text{FD}}(\ETA_{u},\ZETA_{3}) & =\frac{\rho_{d}\eta_{3,k}(M-K)\sigma_{3,k}^{2}}{1+\rho_{d}(\bar{\beta}_{k}-\sigma_{3,k}^2)\sum_{j\in\K}\zeta_{3,j}+\rho_{u}\sum_{i\in\LL}\eta_{u,i}\beta_{\mathrm{IGI},ki}}.\label{eq:SINR_NFL_S3-1-1}
\end{align}
Now, the achievable downlink rate for non-FL UE $k,\forall k\in\K,$
is 
\begin{align}
\!\!\!\!R_{3,k}^{\text{FD}}(\ETA_{u},\ZETA_{3}) & \!=\!\frac{\tau_{c}\!-\!\tau_{3,p}}{\tau_{c}}B\log_{2}\big(1\!+\!\text{SINR}_{3,k}^{\text{FD}}(\ETA_{u},\ZETA_{3})\big).
\end{align}

\textbf{Uplink delay}: Denote by $S_{u}$ (bits) the data size of
the local training update of the FL group. The transmission time from
FL UE $\ell$ to the BS is the same and given by 
\begin{equation}
t_{u}^{\text{FD}}(\ETA_{u},\ZETA_{3})=\frac{S_{u}}{R_{u}^{\text{FD}}(\ETA_{u},\ZETA_{3})}.\label{eq:TimeFLStep3-1}
\end{equation}
The above equation is similar to \eqref{eq:TimeFLStep3} except that
the transmission time now depends on power control coefficients from
both FL and non-FL UEs.

\textbf{Amount of downlink data received at the non-FL group}: The
amount of downlink data received at all non-FL UE $k,\forall k\in\K$,
in Step (S3) is 
\begin{equation}
D_{3,k}^{\text{FD}}(\ETA_{u},\ZETA_{3})=R_{3,k}^{\text{FD}}(\ETA_{u},\ZETA_{3})t_{u}^{\text{FD}}(\ETA_{u},\ZETA_{3}).\label{eq:DataNonFLStep3-1}
\end{equation}
This equation is also similar to \eqref{eq:DataNonFLStep3} while
the only difference is that the downlink rate of $k$-th non-FL UE
and the transmission time of the FL UEs depend on both $\ETA_{u}$
and $\ZETA_{3}$.

\subsubsection{Step (S4)}

After receiving all the local update, the BS (i.e., central server)
computes its global update. since the computational capability of
the central server is much more powerful than those of the UEs, the
delay of computing the global update is negligible.

\section{Problem Formulation and Proposed Solution}

\label{sec:PF}The problem of fairness among the non-FL UEs in terms
of effective data received is one of the key challenges in wireless
communications. 
In this section we first define a new performance metric which is referred to as the effective data rate of non-FL UEs and then formulate the optimization problems to achieve the max-min fairness of non-FL UEs subject to a QoS constraint on the execution time of FL UEs.

\subsection{Effective data rate of non-FL UEs}
From the discussions in the preceding section, the data rate of each non-FL UE is changed for different steps. Thus, it is practically reasonable to use the average data rate accounting for all steps as a representative data rate for the system design purposes. More specifically, the total amount of data received by the $k$-th non-FL UE  in Steps (S1)-(S3) is $D_{1,k}\!+\!D_{2,k}\!+\!D_{3,k}^{\text{mode}}$, where $\text{mode}\in\{\text{HD},\text{FD} \}$. Also, the time of each step is determined by the FL UEs. It is obvious that the total time of the three steps is $t_{d}+t_{C}+t_{u}^{\text{mode}}$. Thus, we define the effective data rate for the $k$-th non-FL UE as $\tfrac{D_{1,k}\!+\!D_{2,k}\!+\!D_{3,k}^{\text{mode}}}{t_{d}+t_{C}+t_{u}^{\text{mode}}}$. In the following we use this definition of the effective data rate for non-FL UEs to formulate max-min fairness problems for HD and FD approaches.
\subsection{HD Scheme}

\subsubsection{Problem Formulation for HD Scheme}

The considered problem for the HD communication scheme can be mathematically
expressed as follows: 
\begin{subequations}
\label{Pmain:HD} 
\begin{align}
\!\!\!\!\!\!\!\!\underset{\x^\text{HD}}{\max}\,\, & \frac{\min_{k\in\K}\big(D_{1,k}(\ETA_{d},\ZETA_{1})\!+\!D_{2,k}(f,\ZETA_{2})\!+\!D_{3,k}^{\text{HD}}(\ETA_{u},\ZETA_{3})\big)}{t_{d}(\ETA_{d},\ZETA_{1})+t_{C}(f)+t_{u}^{\text{HD}}(\ETA_{u})}\\
\mathrm{s.t.}\,\, & \eqref{powerdupperbound},\eqref{powerdupperbound-1},\eqref{poweruupperbound},\eqref{powerdupperbound-2},\nonumber\\
 & \eta_{d,\ell},\zeta_{1,k},\zeta_{2,k},\eta_{u,\ell},\zeta_{3,k}\geq0,\eta_{d,\ell}\leq1,\label{S1powerUbound}\\
 & f_{\min}\leq f_{\ell}\leq f_{\max},\forall\ell\label{fbound}\\
 & t_{d}(\ETA_{d},\ZETA_{1})+t_{C}(f)+t_{u}^{\text{HD}}(\ETA_{u})\leq t_{\text{QoS}}^{\text{HD}},\label{eq:QoSbound:HD}
\end{align}
\end{subequations}
 where $\x^\text{HD}\triangleq\{\ETA_{d},\ZETA_{1},f,\ZETA_{2},\ETA_{u},\ZETA_{3}\}$.
The constraint \eqref{eq:QoSbound:HD} is introduced to ensure that the
time taken by the FL UEs is bounded by $t_{\text{QoS}}^{\text{HD}}$.

\subsubsection{Solution for HD Scheme}

In this section, we present a solution to \eqref{Pmain:HD} based on
successive convex approximation (SCA). Our idea is to equivalent transform the sophisticated constraints into simpler ones where convex approximations are easier to find.  To this end, using the epigraph form, we first equivalently rewrite \eqref{Pmain:HD} as 
\begin{subequations}
\label{Pmain:HD-1} 
\begin{align}
\!\!\!\!\!\!\!\!\underset{\bar{\x}^\text{HD}}{\max}\,\, & \frac{t^{\text{HD}}}{t_{\text{Q}}^{\text{HD}}}\\
\mathrm{s.t.}\,\, & \eqref{powerdupperbound},\eqref{powerdupperbound-1},\eqref{poweruupperbound},\eqref{powerdupperbound-2},\eqref{S1powerUbound},\eqref{fbound},\nonumber\\
 & t_{d}(\ETA_{d},\ZETA_{1})+t_{C}(f)+t_{u}^{\text{HD}}(\ETA_{u})\leq t_{\text{Q}}^{\text{HD}}, \label{eq:QoSbound:HD-1}\\
 & D_{1,k}(\ETA_{d},\ZETA_{1})\!+\!D_{2,k}(f,\ZETA_{2})\!+\!D_{3,k}^{\text{HD}}(\ETA_{u},\ZETA_{3})\geq t^{\text{HD}}, \forall k \label{eq:data_const}\\
 & t_{\text{Q}}^{\text{HD}}\leq t_{\text{QoS}}^{\text{HD}},\label{eq:QoSbound:HD-2}
\end{align}
\end{subequations}
 where $\bar{\x}^\text{HD}=\{\x^{\text{HD}},t^{\text{HD}},t_{\text{Q}}^{\text{HD}}\}$. Next, it is straightforward to see that  \eqref{Pmain:HD-1} can be further equivalently reformulated as 
\begin{subequations}
\label{Pmain:HD-2} 
\begin{align}
\!\!\!\!\!\underset{\bar{\x}^\text{HD}}{\max}\,\, & \frac{t^{\text{HD}}}{t_{\text{Q}}^{\text{HD}}}\\
\mathrm{s.t.}\,\, & \eqref{powerdupperbound},\eqref{powerdupperbound-1},\eqref{poweruupperbound},\eqref{powerdupperbound-2},\eqref{S1powerUbound},\eqref{fbound},\eqref{eq:QoSbound:HD-2},\nonumber\\
 & \frac{S_{d}}{R_{d}(\ETA_{d},\ZETA_{1})}+\frac{N_{c}D_{\max}c_{\max}}{f}+\frac{S_{u}}{R_{u}^{\text{HD}}(\ETA_{u})}\leq t_{\text{Q}}^{\text{HD}},\label{eq:QoSbound:HD-3}\\
 & R_{1,k}(\ETA_{d},\ZETA_{1})\frac{S_{d}}{R_{d}(\ETA_{d},\ZETA_{1})}+R_{2,k}(\ZETA_{2})\frac{N_{c}D_{\max}c_{\max}}{f}+R_{3,k}^{\text{HD}}(\ZETA_{3})\frac{S_{u}}{R_{u}^{\text{HD}}(\ETA_{u})}\geq t^{\text{HD}},\forall k.\label{eq:data_const-2}
\end{align}
\end{subequations}
It is now clear that \eqref{eq:QoSbound:HD-3} and   \eqref{eq:data_const-2} are troublesome. We note that \eqref{eq:QoSbound:HD-3} is equivalent to the following set of constraints
\begin{subequations}\label{eq:tQoS-1}\begin{gather}
\frac{S_{d}}{r_{d}}+\frac{N_{c}D_{\max}c_{\max}}{f}+\frac{S_{u}}{r_{u}^{\text{HD}}}\leq t_{\text{Q}}^\text{HD},\label{eq:tQoSHD:convex}\\
r_{d}  \leq R_{d,\ell}(\ETA_{d},\ZETA_{1}),\forall\ell\label{Rdell-lowerbound}\\
r_{u}^{\text{HD}}  \leq R_{u,\ell}^{\text{HD}}(\ETA_{u}),\forall\ell\label{Ruell-lowerbound}\\
r_{d} \geq 0,r_{u}^{\text{HD}} \geq 0. \label{eq:rdru:pos}
\end{gather}\end{subequations}

 It is easy to see that \eqref{eq:QoSbound:HD-3} and \eqref{eq:tQoS-1} are equivalent since if $R_{d,\ell}(\ETA_{d},\ZETA_{1})$ and $R_{u,\ell}^{\text{HD}}(\ETA_{u})$ are feasible to \eqref{eq:QoSbound:HD-3}, then they are also feasible to \eqref{eq:tQoS-1} and vice versa. We note that \eqref{eq:tQoSHD:convex} is convex. Intuitively, $r_{d}$ and $r_{u}^{\text{HD}}$ are lower-bounds of $R_{d,\ell}(\ETA_{d},\ZETA_{1})$
and $R_{u,\ell}^{\text{HD}}(\ETA_{u})$, respectively, for all $\ell\in\LL$.
In the same way, to deal with \eqref{eq:data_const-2}, we rewrite it as
\begin{subequations}\begin{gather}
r_{1,k}\frac{S_{d}}{\tilde{r}_{d}}+r_{2,k}\frac{N_{c}D_{\max}c_{\max}}{f}+r_{3,k}^{\text{HD}}\frac{S_{u}}{\tilde{r}_{u}^{\text{HD}}}\geq t^{\text{HD}},\forall k \label{eq:QoSbound:HD-4}\\
R_{d,\ell}(\ETA_{d},\ZETA_{1})  \leq\tilde{r}_{d},\forall\ell \label{Rdell-upperbound}\\
R_{u,\ell}^{\text{HD}}(\ETA_{u})  \leq\tilde{r}_{u}^{\text{HD}},\forall\ell,\label{Ruell-upperbound}\\
r_{1,k}\leq  R_{1,k}(\ETA_{d},\ZETA_{1}),\forall k\label{R1k-lowerbound}\\
r_{2,k}\leq  R_{2,k}(\ZETA_{2}),\forall k\label{R2k-lowerbound}\\
r_{3,k}^{\text{HD}}\leq R_{3,k}^{\text{HD}}(\ZETA_{3}),\forall k\label{R3k-lowerbound}\\
\tilde{r}_{d}\geq 0, \tilde{r}_{u}\geq 0, r_{1,k}\geq 0, r_{2,k}\geq 0, r_{3,k}^{\text{HD}} \geq 0, \forall k \label{eq:r123:pos}
\end{gather}
\end{subequations}
where  $\tilde{r}_{d}$
and $\tilde{r}_{u}^{\text{HD}}$ are respectively  seen as upper-bounds of $R_{d,\ell}(\ETA_{d},\ZETA_{1})$
and $R_{u,\ell}^{\text{HD}}(\ETA_{u})$ for all $\ell\in\LL$, and 
$r_{1,k}$, $r_{2,k}$ and $r_{3,k}^{\text{HD}}$ the lower-bounds of $R_{1,k}(\ETA_{d},\ZETA_{1})$,
$R_{2,k}(\ZETA_{2})$ and $R_{3,k}^{\text{HD}}(\ZETA_{3})$. 
We can now further equivalently express \eqref{eq:QoSbound:HD-4} as 
\begin{subequations}\label{eq:tQoS:rewrite-4}
\begin{align}
t^{\text{HD}}&\leq a_{1}S_{d}+a_{2}N_{c}D_{\max}c_{\max}+a_{3}^{\text{HD}}S_{u},\label{eq:QoSbound:HD-4-convex}\\
a_{1} & \leq\frac{r_{1,k}}{\tilde{r}_{d}} \Leftrightarrow a_{1}\tilde{r}_{d}\leq r_{1,k},\forall k\label{ratioS1-lowerbound}\\
a_{2} & \leq\frac{r_{2,k}}{f}\Leftrightarrow a_{2} f \leq r_{2,k},\forall k\label{ratioS2-lowerbound}\\
a_{3}^{\text{HD}} & \leq\frac{r_{3,k}^{\text{HD}}}{\tilde{r}_{u}^{\text{HD}}} \Leftrightarrow a_{3}^{\text{HD}} \tilde{r}_{u}^{\text{HD}} \leq r_{3,k}^{\text{HD}},\forall k\label{ratioS3-lowerbound}.
\end{align}
\end{subequations}
The above transformations means 
\eqref{Pmain:HD} is equivalent to the following problem
\begin{subequations}
\label{Pmain:HD-3} 
\begin{align}
\!\!\!\!\!\underset{\tilde{\x}^{\text{HD}}}{\max}\,\, & z^\text{HD} \\
\mathrm{s.t.}\,\, & z^\text{HD} t_{\text{Q}}^{\text{HD}} \leq t^{\text{HD}}, \label{z}\\
&\eqref{powerdupperbound},\eqref{powerdupperbound-1},\eqref{poweruupperbound},\eqref{powerdupperbound-2},\eqref{S1powerUbound},\eqref{fbound}, \eqref{eq:QoSbound:HD-2},\eqref{eq:tQoS-1},\eqref{Rdell-upperbound}-\eqref{R3k-lowerbound},\eqref{eq:tQoS:rewrite-4},\nonumber
\end{align}
\end{subequations}
 where $\tilde{\x}^{\text{HD}}\triangleq\{\bar{\x}^\text{HD},r_{d},r_{u}^{\text{HD}},a_{1},a_{2},a_{3}^{\text{HD}},\rrr_{1},\rrr_{2},\rrr_{3}^{\text{HD}},\tilde{r}_{d},\tilde{r}_{u}^{\text{HD}},z^\text{HD}\}$,
$\rrr_{1}=\{r_{1,k}\}$, $\rrr_{2}=\{r_{2,k}\}$, $\rrr_{3}^{\text{HD}}=\{r_{3,k}^{\text{HD}}\}$.

Problem \eqref{Pmain:HD-3} is still difficult to solve due to nonconvex constraints  \eqref{Rdell-lowerbound}-\eqref{R3k-lowerbound}, \eqref{ratioS1-lowerbound}-\eqref{ratioS3-lowerbound}, and \eqref{z}. However, these constraints are amenable to applying the SCA method, which we show next.
 
In the sequel, we denote $\tilde{\x}^{\text{HD}(n)}$ to be the value of $\tilde{\x}^{\text{HD}}$ after $n$ iterations.  We first note that constraints \eqref{Rdell-lowerbound}, \eqref{Ruell-lowerbound},
\eqref{R1k-lowerbound}--\eqref{R3k-lowerbound} are of the same type in the sense that concave lower bounds of the involving rate expressions are required to obtain convex approximate constraints. To this end we recall the
following inequality
\begin{align}
\log\big(1+\frac{x}{y}\big)\geq & \log\big(1+\frac{x^{(n)}}{y^{(n)}}\big)+\frac{2x^{(n)}}{(x^{(n)}+y^{(n)})}-\frac{(x^{(n)})^{2}}{(x^{(n)}+y^{(n)})x}-\frac{x^{(n)}y}{(x^{(n)}+y^{(n)})y^{(n)}},\label{eq:logapproxLB}
\end{align}
where $x>0,y>0$ \cite[(76)]{long21TCOM}. Applying the above inequality
we obtain the following inequalities 
\begin{subequations}
\begin{align}
\tilde{R}_{d,\ell}(\ETA_{d},\ZETA_{1}) & \leq  R_{d,\ell}(\ETA_{d},\ZETA_{1}),\forall\ell \label{Rd_tilde_LB}\\
\tilde{R}_{u,\ell}^{\text{HD}}(\ETA_{u}) & \leq R_{u,\ell}^{\text{HD}}(\ETA_{u}),\forall\ell\label{Ru_tilde_LB}\\
\tilde{R}_{1,k}(\ETA_{d},\ZETA_{1}) & \leq R_{1,k}(\ETA_{d},\ZETA_{1}),\forall k\label{R1k_tilde_LB}\\
\tilde{R}_{2,k}(\ZETA_{2}) & \leq R_{2,k}(\ZETA_{2}),\forall k\label{R2k_tilde_LB}\\
\tilde{R}_{3,k}^{\text{HD}}(\ZETA_{3}) & \leq R_{3,k}^{\text{HD}}(\ZETA_{3}),\forall k\label{R3k_tilde_LB}
\end{align}
\end{subequations}
where $\tilde{R}_{d,\ell}(\ETA_{d},\ZETA_{1})$, $\tilde{R}_{u,\ell}^{\text{HD}}(\ETA_{u})$,
$\tilde{R}_{1,k}(\ETA_{d},\ZETA_{1})$, $\tilde{R}_{2,k}(\ZETA_{2})$,
and $\tilde{R}_{3,k}^{\text{HD}}(\ZETA_{3})$ are concave lower
bounds of $R_{d,\ell}(\ETA_{d},\ZETA_{1})$, $R_{u,\ell}^{\text{HD}}(\ETA_{u})$,
$R_{1,k}(\ETA_{d},\ZETA_{1})$, $R_{2,k}(\ZETA_{2})$, and $R_{3,k}^{\text{HD}}(\ZETA_{3})$,
respectively. The expressions of these lower bounds are given in \eqref{eq:rate:HD:LB} in Appendix \ref{sec:LBrates}.
Consequently, in light of SCA, \eqref{Rdell-lowerbound}, \eqref{Ruell-lowerbound},
\eqref{R1k-lowerbound}--\eqref{R3k-lowerbound} are approximated
by the following convex constraints 
\begin{subequations}\label{eq:LBtype}
\begin{align}
r_{d}\leq & \tilde{R}_{d,\ell}(\ETA_{d},\ZETA_{1}),\forall\ell\label{Rdell-lowerbound:approx}\\
r_{u}^{\text{HD}}\leq & \tilde{R}_{u,\ell}^{\text{HD}}(\ETA_{u}),\forall\ell\label{Ruell-lowerbound:approx}\\
r_{1,k}\leq & \tilde{R}_{1,k}(\ETA_{d},\ZETA_{1}),\forall k\label{R1k-lowerbound:approx}\\
r_{2,k}\leq & \tilde{R}_{2,k}(\ZETA_{2}),\forall k\label{R2k-lowerbound:approx}\\
r_{3,k}^{\text{HD}}\leq & \tilde{R}_{3,k}^{\text{HD}}(\ZETA_{3}),\forall k.\label{R3k-lowerbound:approx}
\end{align}
\end{subequations}
To proceed further we note that constraints \eqref{ratioS1-lowerbound}--\eqref{ratioS3-lowerbound} and \eqref{z} are of the same type.
To deal with these, let us recall the following equality
\begin{align}
xy = \frac{1}{4} [(x+y)^2-(x-y)^2].\label{xy}
\end{align}
Since we need a \emph{convex upper bound} of the term $xy$, a simple way is to linearize the term  
$(x-y)^{2}$.
In this way we arrive at the following inequality
\begin{equation}
xy \leq\frac{1}{4} [(x+y)^2-2(x^{(n)}-y^{(n)})(x-y)
+ (x^{(n)}-y^{(n)})^2],\label{quadraticapprox}
\end{equation}
where $x\geq0, y\geq0$, and $x^{(n)}$ and $y^{(n)}$ are the values of $x$ and $y$ at
the $n$-th iteration, respectively \cite{vu20TWC}. Thus, using \eqref{quadraticapprox} we can approximate \eqref{ratioS1-lowerbound}--\eqref{ratioS3-lowerbound} and \eqref{z} by the following convex constraints, respectively
\begin{subequations}\label{eq:xy:UB}
\begin{gather}
\frac{1}{4}[(a_{1}+\tilde{r}_{d})^{2}-(a_{1}^{(n)}-\tilde{r}_{d}^{(n)})^{2}+2(a_{1}-\tilde{r}_{d})(a_{1}^{(n)}-\tilde{r}_{d}^{(n)})]\leq r_{1,k},\forall k\label{ratioS1-lowerbound:approx}\\
\frac{1}{4}[(a_{2}+f)^{2}-(a_{2}^{(n)}-f^{(n)})^{2}+2(a_{2}-f)(a_{2}^{(n)}-f^{(n)})] \leq r_{2,k},\forall k\label{ratioS2-lowerbound:approx}\\
\frac{1}{4}[(a_{3}^{\text{HD}}+\tilde{r}_{u}^{\text{HD}})^{2}-(a_{3}^{\text{HD}(n)}-\tilde{r}_{u}^{\text{HD}(n)})^{2}+2(a_{3}^{\text{HD}}-\tilde{r}_{u}^{\text{HD}})(a_{3}^{\text{HD}(n)}-\tilde{r}_{u}^{\text{HD}(n)})]\leq r_{3,k}^{\text{HD}},\forall k\label{ratioS3-lowerbound:approx}\\
\frac{1}{4}\big[(z^{\text{HD}}+t_\text{Q}^{\text{HD}})^{2}-(z^{\text{HD}(n)}-t_\text{Q}^{\text{HD}(n)})^{2}+2(z^{\text{HD}}-t_\text{Q}^{\text{HD}})(z^{\text{HD}(n)}-t_\text{Q}^{\text{HD}(n)}\big)\big]\leq t^\text{HD}.\label{z:approx}
\end{gather}
\end{subequations}

We now turn our attention to \eqref{Rdell-upperbound} and \eqref{Ruell-upperbound}. It is obvious now we need to derive \emph{convex upper bounds} of the rate functions present in these two constraints. To this end we resort to the following inequality
\begin{equation}
\log\big(1+\frac{x}{y}\big)\leq \log\big(1+\frac{x^{(n)}}{y^{(n)}}\big)+\frac{y^{(n)}}{(x^{(n)}+y^{(n)})}\times\big(\frac{(x^{2}+(x^{(n)})^{2})}{2x^{(n)}y}-\frac{x^{(n)}}{y^{(n)}}\big),\label{eq:logapproxUB}
\end{equation}
where $x>0,y>0$, and $x^{(n)}$ and $y^{(n)}$ are the values of
$x$ and $y$ at the $n$-th iteration, respectively\cite[(75)]{sheng18TWC}.
Using this inequality we can approximate constraints \eqref{Rdell-upperbound} and \eqref{Ruell-upperbound}
by the following convex constraints 
\begin{subequations}\label{eq:rate:LB}
\begin{align}
\hat{R}_{d,\ell}(\ETA_{d},\ZETA_{1}) & \leq\tilde{r}_{d},\forall\ell\label{Rdell-upperbound:approx}\\
\hat{R}_{u,\ell}^{\text{HD}}(\ETA_{u}) & \leq\tilde{r}_{u}^{\text{HD}},\forall\ell\label{Ruell-upperbound:approx}
\end{align}
\end{subequations}
where $\hat{R}_{d,\ell}(\ETA_{d},\ZETA_{1})$ and $\hat{R}_{u,\ell}^{\text{HD}}(\ETA_{u})$
are convex upper bounds of ${R}_{d,\ell}(\ETA_{d},\ZETA_{1})$
and ${R}_{u,\ell}^{\text{HD}}(\ETA_{u})$, respectively. The detailed expressions of $\hat{R}_{d,\ell}(\ETA_{d},\ZETA_{1})$ and $\hat{R}_{u,\ell}^{\text{HD}}(\ETA_{u})$ are given in \eqref{eq:rate:HD:UB} in Appendix \ref{sec:LBrates}.

In summary, at iteration $n+1$, problem \eqref{Pmain:HD-2} is approximated by the following convex problem:
\begin{equation}\label{eq:convexprob}
    \max \ \{z^{\text{HD}}\ |\  \tilde{\x}^{\text{HD}} \in\tilde{\F}^\text{HD}\},
\end{equation}
where $\tilde{\F}^\text{HD}\triangleq \{\eqref{powerdupperbound},\eqref{powerdupperbound-1},\eqref{poweruupperbound},\eqref{powerdupperbound-2},\eqref{S1powerUbound},\eqref{fbound},\eqref{eq:QoSbound:HD-2},\eqref{eq:tQoSHD:convex},   
\eqref{eq:rdru:pos}, \eqref{eq:r123:pos}, \eqref{eq:QoSbound:HD-4-convex}, \eqref{eq:LBtype},\eqref{eq:xy:UB},\eqref{eq:rate:LB}\}$. We outline the main steps to solve problem \eqref{Pmain:HD-3} in Algorithm~\ref{alg:SCA}.
\begin{algorithm}[t!]
	\caption{Algorithm for solving \eqref{Pmain:HD}}
	\label{alg:SCA}
	\begin{algorithmic}[1]
		\STATE Input: Set $n=0$ and choose an initial point $\tilde{\x}^{\text{HD}(0)}\in\tilde{\F}^\text{HD}$
		\REPEAT
		\STATE Solve \eqref{eq:convexprob} to get $\tilde{\x}^{\text{HD}\ast}$
		\STATE $\tilde{\x}^{\text{HD}(n+1)}\leftarrow \tilde{\x}^{\text{HD}\ast}$
		\STATE $n\leftarrow n+1$
		\UNTIL{convergence}
	\end{algorithmic}
\end{algorithm}
\begin{rem}\label{rem1} Algorithm~\ref{alg:SCA} requires a feasible point to start the iterative procedure. In general, it is difficult to find a feasible solution to \eqref{Pmain:HD-3}. We now describe a practical way to overcome this issue. 
It is not difficult to see that by randomly generating and properly the variables in $\x^\text{HD}$ we can meet \eqref{powerdupperbound}, \eqref{powerdupperbound-1}, \eqref{poweruupperbound}, \eqref{powerdupperbound-2}, \eqref{S1powerUbound}, \eqref{fbound}. The remaining variables in $\tilde{\x}^{\text{HD}}$ can be found by letting the corresponding inequality constraint \eqref{Pmain:HD-3}  be binding (i.e., occur with equality). If \eqref{eq:QoSbound:HD-2} is satisfied, then we can use this initial point to start Algorithm ~\ref{alg:SCA}. When the requirements are high (e.g., when $t_{\text{QoS}}^{\text{HD}}$ is small), it is likely that \eqref{eq:QoSbound:HD-2} is not met. In such cases, we introduce a slack variable $s$ and replacing \eqref{eq:convexprob} by the following problem 
\begin{subequations}\label{eq:subproblem:relax}
\begin{align}
\underset{s\leq 0, \tilde{\x}^\text{HD}}{\max}\,\, & z^\text{HD}+\alpha s\label{eq:subproblem:relax:obj}\\
\mathrm{s.t.}\,\,&  \tilde{\x}^\text{HD} \in \tilde{\F}^{\text{HD}} \setminus \eqref{eq:QoSbound:HD-2},\\
\,\,&t_{\text{Q}}^{\text{HD}}+s\leq t_{\text{QoS}}^{\text{HD}}\label{eq:tQos:relax}.
\end{align}
\end{subequations}
\end{rem}
Intuitively, $s$ represents the violation of \eqref{eq:QoSbound:HD-2} and $\alpha>0 $ is the penalty parameter. It is easy to see that \eqref{eq:tQos:relax} is met if $s$ is sufficiently small, and thus  \eqref{eq:subproblem:relax} is always feasible. On the other hand, the maximization of the regularized objective in \eqref{eq:subproblem:relax:obj} will force $s$ to approach $0$ when the iterative process progresses. Thus, when Algorithm~\ref{alg:SCA}  converges and if $|s|$ is smaller than a pre-determined error tolerance, we will take $\tilde{\x}^{\text{HD}\ast}$ as the final solution. Otherwise, we say that \eqref{Pmain:HD} is infeasible. 

\subsection{FD Scheme}

\subsubsection{Problem Formulation for FD Communication Scheme}

The considered problem for the FD communication scheme is mathematically
stated as 
\begin{subequations}
\label{Pmain:FD} 
\begin{align}
\!\!\!\!\!\!\!\!\underset{\x^\text{FD}}{\max}\,\, & \frac{\min_{k\in\K}\big(D_{1,k}(\ETA_{d},\ZETA_{1})\!+\!D_{2,k}(f,\ZETA_{2})\!+\!D_{3,k}^{\text{FD}}(\ETA_{u},\ZETA_{3})\big)}{t_{d}(\ETA_{d},\ZETA_{1})+t_{C}(f)+t_{u}^{\text{FD}}(\ETA_{u},\ZETA_{3})}\\
\mathrm{s.t.}\,\, & \eqref{powerdupperbound},\eqref{powerdupperbound-1},\eqref{poweruupperbound},\eqref{powerdupperbound-2},\nonumber\\
 & \eta_{d,\ell},\zeta_{1,k},\zeta_{2,k},\eta_{u,\ell},\zeta_{3,k}\geq0,\eta_{d,\ell}\leq1,\label{S1powerUbound2}\\
 & f_{\min}\leq f_{\ell}\leq f_{\max},\forall\ell\label{fbound2}\\
 & t_{d}(\ETA_{d},\ZETA_{1})+t_{C}(f)+t_{u}^{\text{FD}}(\ETA_{u},\ZETA_{3})\leq t_{\text{QoS}}^{\text{FD}},\label{eq:QoSbound:FD}
\end{align}
\end{subequations}
where $\x^{\text{FD}} \triangleq\{\ETA_{d},\ZETA_{1},f,\ZETA_{2},\ETA_{u},\ZETA_{3}\}$.

\subsubsection{Solution for FD Scheme}
The solution for the FD scheme follows closely the derivations of that for the HD scheme.  First, we equivalently rewrite \eqref{Pmain:FD} as 
\begin{subequations}
\label{Pmain:FD-1} 
\begin{align}
\!\!\!\!\!\underset{\bar{\x}^\text{FD}}{\max}\,\, & \frac{t^{\text{FD}}}{t_{\text{Q}}^{\text{FD}}}\\
\mathrm{s.t.}\,\, & \eqref{powerdupperbound},\eqref{powerdupperbound-1},\eqref{poweruupperbound},\eqref{powerdupperbound-2},\eqref{S1powerUbound2},\eqref{fbound2},\nonumber\\
 & R_{1,k}(\ETA_{d},\ZETA_{1})\frac{S_{d}}{R_{d}(\ETA_{d},\ZETA_{1})}+R_{2,k}(\ZETA_{2})\frac{N_{c}D_{\max}c_{\max}}{f}+R_{3,k}^{\text{FD}}(\ETA_{u},\ZETA_{3})\frac{S_{u}}{R_{u}^{\text{FD}}(\ETA_{u},\ZETA_{3})}\geq t,\forall k\label{eq:data_const2}\\
 & \frac{S_{d}}{R_{d}(\ETA_{d},\ZETA_{1})}+\frac{N_{c}D_{\max}c_{\max}}{f}+\frac{S_{u}}{R_{u}^{\text{FD}}(\ETA_{u},\ZETA_{3})}\leq t_{\text{Q}}^{\text{FD}},\label{eq:QoSbound:FD-1}\\
 & t_{\text{Q}}^{\text{FD}}\leq t_{\text{QoS}}^{\text{FD}},\label{eq:QoSbound:FD-2}
\end{align}
\end{subequations}
where $\bar{\x}^\text{FD}=\{\x^{\text{FD}},t^{\text{FD}},t_{\text{Q}}^{\text{FD}}\}$, which is then equivalent to
\begin{subequations}
\label{Pmain:FD-2} 
\begin{align}
\!\!\!\!\!\underset{\tilde{\x}^{\text{FD}}}{\max}\,\, & z^\text{FD} \\
\mathrm{s.t.}\,\, & \eqref{powerdupperbound},\eqref{powerdupperbound-1},\eqref{poweruupperbound},\eqref{powerdupperbound-2},\eqref{S1powerUbound2},\eqref{fbound2},\eqref{eq:QoSbound:FD-2},\nonumber \\
 & a_{1}S_{d}+a_{2}N_{c}D_{\max}c_{\max}+a_{3}^{\text{FD}}S_{u}\geq t^{\text{FD}},\forall k\label{CFPmain-lowerbound-1}\\
 & \frac{S_{d}}{r_{d}}+\frac{N_{c}D_{\max}c_{\max}}{f}+\frac{S_{u}}{r_{u}^{\text{FD}}}\leq t_{\text{Q}}^{\text{FD}},\label{eq:QoSbound:FD-3}\\
 & z^{\text{FD}}t_{\text{Q}}^{\text{FD}}\leq t^{\text{FD}},\label{z_FD}\\
 & a_{1}\tilde{r}_{d}\leq {r_{1,k}},\forall k\label{ratioS1-lowerbound-1}\\
 & a_{2} f \leq{r_{2,k}},\forall k\label{ratioS2-lowerbound-1}\\
 & a_{3}^{\text{FD}}\tilde{r}_{u}^{\text{FD}}\leq {r_{3,k}^{\text{FD}}},\forall k\label{ratioS3-lowerbound-1}\\
 & r_{d}\leq R_{d,\ell}(\ETA_{d},\ZETA_{1}),\forall\ell\label{Rdell-lowerbound-1}\\
 & r_{u}^{\text{FD}}\leq R_{u,\ell}^{\text{FD}}(\ETA_{u},\ZETA_{3}),\forall\ell\label{Ruell-lowerbound-1}\\
 & r_{1,k}\leq R_{1,k}(\ETA_{d},\ZETA_{1}),\forall k\label{R1k-lowerbound-1}\\
 & r_{2,k}\leq R_{2,k}(\ZETA_{2}),\forall k\label{R2k-lowerbound-1}\\
 & r_{3,k}^{\text{FD}}\leq R_{3,k}^{\text{FD}}(\ETA_{u},\ZETA_{3}),\forall k\label{R3k-lowerbound-1}\\
 & R_{d,\ell}(\ETA_{d},\ZETA_{1})\leq\tilde{r}_{d},\forall\ell\label{Rdell-upperbound-1}\\
 & R_{u,\ell}^{\text{FD}}(\ETA_{u},\ZETA_{3})\leq\tilde{r}_{u}^{\text{FD}},\forall\ell\label{Ruell-upperbound-1}
\end{align}
\end{subequations}
 where $\tilde{\x}^{\text{FD}}\triangleq\{\bar{\x},r_{d},r_{u}^{\text{FD}},a_{1},a_{2},a_{3}^{\text{FD}},\rrr_{1},\rrr_{2},\rrr_{3}^{\text{FD}},\tilde{r}_{d},\tilde{r}_{u}^{\text{FD}},z^\text{FD}\}$,
$\rrr_{1}=\{r_{1,k}\}$, $\rrr_{2}=\{r_{2,k}\}$, $\rrr_{3}^{\text{FD}}=\{r_{3,k}^{\text{FD}}\}$. It is clear that the nonconvexity of problem \eqref{Pmain:FD-2} is due to \eqref{z_FD}-\eqref{Ruell-upperbound-1}. 
We remark that \eqref{ratioS1-lowerbound-1} and \eqref{ratioS2-lowerbound-1} are indeed \eqref{ratioS1-lowerbound} and \eqref{ratioS2-lowerbound}, respectively, and their convex approximations are given in \eqref{ratioS1-lowerbound:approx} and \eqref{ratioS2-lowerbound:approx}. 
Similar to \eqref{ratioS3-lowerbound:approx} and \eqref{z:approx},  constraints  \eqref{z_FD} and \eqref{ratioS3-lowerbound-1} can be approximated by the following convex constraints
\begin{subequations}\label{eq:tFD}
\begin{align}
\frac{1}{4}\big[(z^{\text{HD}}+t_\text{Q}^{\text{HD}})^{2}-(z^{\text{HD}(n)}-t_\text{Q}^{\text{HD}(n)})^{2}+2(z^{\text{HD}}-t_\text{Q}^{\text{HD}})(z^{\text{HD}(n)}-t_\text{Q}^{\text{HD}(n)}\big)\big] &\leq t^\text{HD},\label{z_FD:approx}\\
\frac{1}{4}[(a_{3}^{\text{FD}}+\tilde{r}_{u}^{\text{FD}})^{2}-(a_{3}^{\text{FD}(n)}-\tilde{r}_{u}^{\text{FD}(n)})^{2}+2(a_{3}^{\text{FD}}-\tilde{r}_{u}^{\text{FD}})(a_{3}^{\text{FD}(n)}-\tilde{r}_{u}^{\text{FD}(n)})]&\leq r_{3,k}^{\text{FD}},\forall k\label{ratioS3-lowerbound:approx-1}
\end{align}
\end{subequations}
where $a_{3}^{\text{FD}(n)}$ and $\tilde{r}_{u}^{\text{FD}(n)}$
are the values of $a_{3}^{\text{FD}}$ and $\tilde{r}_{u}^{\text{FD}}$
at the $n$-th iteration, respectively.

To proceed further, we note that the convex approximate constraints of \eqref{Rdell-lowerbound-1}, \eqref{R1k-lowerbound-1},  \eqref{R2k-lowerbound-1}, and \eqref{Rdell-upperbound-1} are already presented in \eqref{Rdell-lowerbound:approx}, \eqref{R1k-lowerbound:approx}, \eqref{R2k-lowerbound:approx}, and \eqref{Rdell-upperbound:approx}, respectively.
Also, \eqref{Ruell-lowerbound-1}, \eqref{R3k-lowerbound-1}, and \eqref{Ruell-upperbound-1} are similar to \eqref{Ruell-lowerbound}, \eqref{R3k-lowerbound}, and \eqref{Ruell-upperbound}. Thus, following the same steps to obtaining \eqref{Ruell-lowerbound:approx}, \eqref{R3k-lowerbound:approx}, and \eqref{Ruell-upperbound:approx},  we can approximate \eqref{Ruell-lowerbound-1}, \eqref{R3k-lowerbound-1}, and \eqref{Ruell-upperbound-1} as the following convex constraints
\begin{subequations}\label{eq:rateUB:FD}
\begin{align}
r_{u}^{\text{FD}} & \leq\tilde{R}_{u,\ell}^{\text{FD}}(\ETA_{u},\ZETA_{3}),\forall\ell\label{Ruell-lowerbound:approx-1}\\
r_{3,k}^{\text{FD}} & \leq\tilde{R}_{3,k}^{\text{FD}}(\ETA_{u},\ZETA_{3}),\forall k\label{R3k-lowerbound:approx-1}\\
\hat{R}_{u,\ell}^{\text{FD}}(\ETA_{u},\ZETA_{3}) & \leq\tilde{r}_{u}^{\text{FD}},\forall\ell\label{Ruell-upperbound:approx-1}
\end{align}
\end{subequations}
where $\tilde{R}_{u,\ell}^{\text{FD}}(\ETA_{u},\ZETA_{3})$,  $\tilde{R}_{3,k}^{\text{FD}}(\ETA_{u},\ZETA_{3})$, and $\hat{R}_{u,\ell}^{\text{FD}}(\ETA_{u},\ZETA_{3})$ are given in \eqref{eq:rate:FD:LB} and \eqref{eq:rate:FD:UB} in Appendix~\ref{sec:LBrates}.

At iteration $n+1$, for a given point $\tilde{\x}^{\text{FD}(n)}$,  problem \eqref{Pmain:FD} is approximated by the following convex problem:
\begin{align}
{\max}\,\, & \{z^\text{FD} \ | \  \tilde{\x}^\text{FD}\in\tilde{\F}^\text{FD}\},\label{eq:convexprob-1}
\end{align}
where $\tilde{\F}^\text{FD}\triangleq \{\eqref{powerdupperbound}, \eqref{powerdupperbound-1}, \eqref{poweruupperbound}, \eqref{powerdupperbound-2}, \eqref{S1powerUbound2}, \eqref{fbound2}, \eqref{eq:QoSbound:FD-2},\eqref{CFPmain-lowerbound-1},\eqref{eq:QoSbound:FD-3}, \eqref{ratioS1-lowerbound:approx}, \eqref{ratioS2-lowerbound:approx},\eqref{Rdell-lowerbound:approx}, \eqref{R1k-lowerbound:approx},\\ \eqref{R2k-lowerbound:approx}, \eqref{Rdell-upperbound:approx}, \eqref{eq:tFD},\eqref{eq:rateUB:FD}\}$. 
We outline the main steps to solve problem \eqref{Pmain:FD-2} in Algorithm~\ref{alg:SCA:FD}.
\begin{rem}\label{rem2} Similar to Algorithm~\ref{alg:SCA}, Algorithm \ref{alg:SCA:FD} requires a feasible point to \eqref{Pmain:FD-2} which is not trivial for find, especially when the SI is high.
To overcome this issue we follow the same procedure as described in Remark \ref{rem1}. Specifically, if scaling randomly generated variables cannot produce a feasible solution, we introduce add a slack variable $s$ and consider the following problem
\begin{subequations}\label{eq:subproblem:FD:relax}
\begin{align}
\underset{s\leq 0,\tilde{\x}^\text{FD}}{\max}\,\, & z^\text{FD}+s\\
\mathrm{s.t.}\,\,& \tilde{\x}^\text{FD} \in \tilde{\F}^\text{FD} \setminus \eqref{eq:QoSbound:FD-2},\\
\,\,&t_{\text{Q}}^{\text{FD}}+s\leq t_{\text{QoS}}^{\text{FD}}.
\end{align}
\end{subequations}
Then, problem \eqref{eq:subproblem:FD:relax} is solved iteratively until convergence. If  $|s|$ is smaller a pre-determined error tolerance, we will take  $\tilde{\x}^\text{FD}$ as the final solution. Otherwise, \eqref{Pmain:FD-2} (and thus \eqref{Pmain:FD}) is said to be infeasible.
\end{rem}
\begin{algorithm}[t!]
	\caption{Algorithm for solving \eqref{Pmain:FD}}
	\label{alg:SCA:FD}
	\begin{algorithmic}[1]
		\STATE Input: Set $n=0$ and choose an initial point $\tilde{\x}^{\text{FD}(0)}\in\tilde{\F}^\text{FD}$
		\REPEAT
		\STATE Solve \eqref{eq:convexprob-1} to get $\tilde{\x}^{\text{FD}\ast}$
		\STATE $\tilde{\x}^{\text{FD}(n+1)}\leftarrow \tilde{\x}^{\text{FD}\ast}$
		\STATE $n\leftarrow n+1$
		\UNTIL{convergence}
	\end{algorithmic}
\end{algorithm}

\section{Numerical Examples}
\subsection{Parameter Setting}
We consider a $D\times D~\text{m}^{2}$ area where the BS is at the centre, while $L$ FL UEs and $K$ non-FL
UEs are randomly distributed. The large-scale fading coefficients are modeled in the same manner as \cite[Eq. (46)]{ziya21TCOM}: 
\begin{align}
     \beta_\ell[\text{dB}] = - 148.1  - 37.6 \log_{10}\big(\frac{d_\ell}{1\,\,\text{km}}\big) + z_\ell,
\end{align}
where $d_\ell \geq 35$ m is the distance between UE $\ell$ and the BS, $z_\ell$ is a shadow fading coefficient which is modeled using a log-normal distribution having zero mean and $7$ dB standard deviation. We set $N_{0}=-92$ dBm, 
$t_{\text{QoS}}=3$ s, $B=20$ MHz, $\rho_{d}=10$ W, $\rho_{u}=\rho_{p}=0.2$
W, $\tau_{d,p}=\tau_{u,p}=20$, $\tau_{S_{1},p} =\tau_{S_{2},p} =\tau_{S_{3},p}=20$, $\tau_{c}=200$, $f_{\min}=0$, $f_{\max}=5\times10^{9}$ cycles/s, $D_{\ell}=D_{\max}=1.6\times10^5$ samples, $c_{\ell}=c_{\max}=20$ cycles/sample, $N_c=20$, $S_d=S_u=16\times10^6$ bits or 16Mb. The path loss $\beta_{\text{SI}}$ is taken as $\beta_{\text{SI}}=10^{\frac{\text{PL}}{10}}$, where $\text{PL}=-81.1846$ dB \cite{vu19ICC}. If not otherwise mentioned, the value of ${\sigma_{\text{SI},0}^2}/{N_0}$ is set to $20$ dB.

\vspace{-5mm}
\subsection{Results and Discussions}
Since there are no other existing works that study massive MIMO networks for supporting both FL and non-FL groups, we compare our proposed scheme with two baseline schemes as follows.
\begin{itemize}
	\item \textbf{BL1}: Steps (S1) and (S3) of this scheme have the same designs as shown in the proposed scheme. In Step (S3), the uplink transmission for the FL group and the downlink of the non-FL group are executed using a  frequency-division multiple access (FDMA) approach for transmission. In particular, we divide the frequency band into all UEs such that each FL UE or non-FL UE has one single bandwidth slot for its transmission. This FDMA scheme is widely used in FL literature (e.g., \cite{Yang2021EEFL,Kim20FDMA}). 
	The uplink and downlink rates of FL UEs in \textbf{BL1} are derived in Appendix \ref{sec:BL1rates}.
	The optimization problem of \textbf{BL1} has the same mathematical structure as that of the proposed scheme. Therefore, it can be solved by slightly modifying Algorithm~\ref{alg:SCA} using the same approximations.
	\item \textbf{BL2}: The downlink powers to FL and non-FL UEs in Step (S1) are equal, i.e., $\eta_{d,\ell}\!=\zeta_{1,k}=\!\frac{1}{L+K}, \forall \ell,k$. The downlink powers to non-FL UEs in Step (S2) and (S3) are also the same, i.e, $\zeta_{2,k}=\zeta_{3,k}=\frac{1}{K}, \forall k$. In addition, in Step (S3), each FL UE uses full power, i.e, $\eta_{u,\ell}=1, \forall \ell$.
	The processing frequencies are $f = \frac{N_cD_mc_{m}}{t_{\text{QoS}-t_{d} - t_{u}}}$. 
\end{itemize}

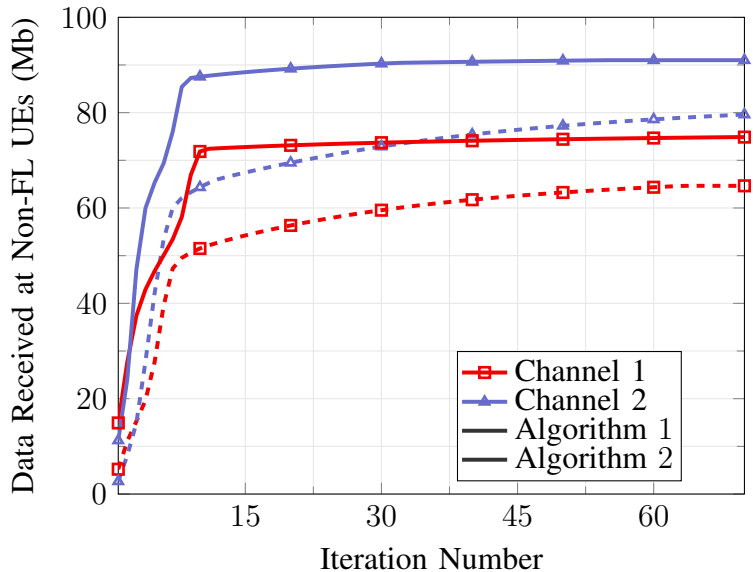
\begin{figure}[t!]
\centering 
\pgfplotsset{width=0.6\columnwidth,height =0.48\columnwidth, compat=1.6}
\setlength{\baselineskip}{0pt}
\begin{tikzpicture}
\begin{axis}[
xtick={15,30,45,60},
ymin=0,ymax=100,xmin=1,xmax=70,
grid=both,
minor tick num = 1,
grid style={mygray},
legend style={anchor=north east,draw=black,fill=white,legend cell align=left,inner sep=1pt,row sep = -3pt,at={(0.9,0.3)}},
xlabel={Iteration Number},
ylabel={Data Received at Non-FL UEs (Mb)},
legend entries={Channel 1, Channel 2, Algorithm $1$, Algorithm $2$}
]
\addlegendimage{mark=square,myred,line width=\lw}
\addlegendimage{mark=triangle,myblue,line width=\lw}
\addlegendimage{mark=dashed,myblack,line width=\lw}
\addlegendimage{mark=none,myblack,line width=\lw}

\addplot[dashed,mark=none, line width=\lw, draw=myred] table [y=HD1, x=iter,col sep = comma]{convergenceJ4.csv};
\addplot[only marks,mark=square, line width=\lw, draw=myred] table [y=HD11, x=iter2,col sep = comma]{convergenceJ4.csv};
\addplot[dashed,mark=none, line width=\lw, draw=myblue] table [y=HD2, x=iter,col sep = comma]{convergenceJ4.csv};
\addplot[only marks,mark=triangle, line width=\lw, draw=myblue] table [y=HD22, x=iter2,col sep = comma]{convergenceJ4.csv};
\addplot[mark=none, line width=\lw, draw=myred] table [y=FD1, x=iter,col sep = comma]{convergenceJ4.csv};
\addplot[only marks,mark=square, line width=\lw, draw=myred] table [y=FD11, x=iter2,col sep = comma]{convergenceJ4.csv};
\addplot[mark=none, line width=\lw, draw=myblue] table [y=FD2, x=iter,col sep = comma]{convergenceJ4.csv};
\addplot[only marks,mark=triangle, line width=\lw, draw=myblue] table [y=FD22, x=iter2,col sep = comma]{convergenceJ4.csv};

%

\end{axis}
\end{tikzpicture} 
\caption{Convergence of the proposed Algorithm $1$ and Algorithm $2$ for two different channel realizations. Here $L=K=5$, and $M=50$.}
\label{convergence} 
\end{figure}

We first provide the convergence of the proposed scheme in comparison with BL1 and BL2 schemes. The convergence plot is shown in Fig. \ref{convergence}. 
It can be observed that  both algorithms converge  in less than 30 iterations for both channel realizations. Further, we note that for both channels, FD-based solution provides a better objective than the HD-based solution.

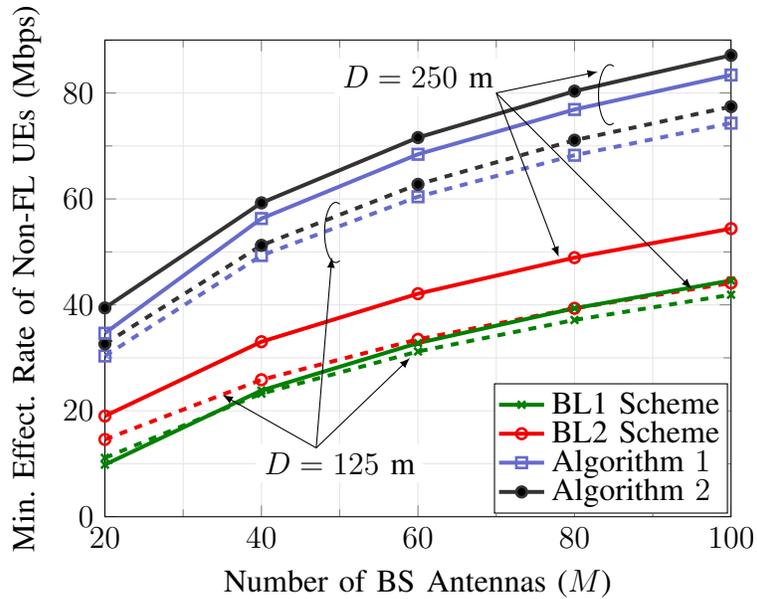
\begin{figure}[t!]
	\centering
	\pgfplotsset{width=0.6\columnwidth,height =0.48\columnwidth, compat=1.6}
	\setlength{\baselineskip}{0pt}
\begin{tikzpicture}
\begin{axis}[
xtick={20,40,60,80,100},
ymin=0,ymax=90,xmin=20,xmax=100,
grid=both,
minor tick num = 1,
grid style={mygray},
legend style={anchor=north east,draw=black,fill=white,legend cell align=left,inner sep=1pt,row sep = -3pt,at={(0.995,0.28)}},
xlabel={Number of BS Antennas ($M$)},
ylabel={Min. Effect. Rate of Non-FL UEs (Mbps)},
legend entries={BL1 Scheme, BL2 Scheme, Algorithm $1$, Algorithm $2$}
]
\addlegendimage{mark=x,line width=\lw, draw=mygreen}
\addlegendimage{mark=o,line width=\lw, draw=myred}
\addlegendimage{mark=square,line width=\lw, draw=myblue}
\addlegendimage{mark=*,line width=\lw, draw=myblack}

\addplot[dashed,mark=o, line width=\lw, draw=myred] table [y=EPA11, x=iter,col sep = comma]{antennaJ4.csv};
\addplot[dashed,mark=square, line width=\lw, draw=myblue] table [y=HD11, x=iter,col sep = comma]{antennaJ4.csv};
\addplot[dashed,mark=*, line width=\lw, draw=myblack] table [y=FD11, x=iter,col sep = comma]{antennaJ4.csv};
\addplot[dashed,mark=x, line width=\lw, draw=mygreen] table [y=BL11, x=iter,col sep = comma]{antennaJ4.csv};
\addplot[mark=o, line width=\lw, draw=myred] table [y=EPA22, x=iter,col sep = comma]{antennaJ4.csv};
\addplot[mark=square, line width=\lw, draw=myblue] table [y=HD22, x=iter,col sep = comma]{antennaJ4.csv};
\addplot[mark=*, line width=\lw, draw=myblack] table [y=FD22, x=iter,col sep = comma]{antennaJ4.csv};
\addplot[mark=x, line width=\lw, draw=mygreen] table [y=BL22, x=iter,col sep = comma]{antennaJ4.csv};

\node[align=center,fill=white,inner sep=3pt] (NoPC) at (axis cs: 60,83) {$D=250$ m};
\draw (axis cs: 85,85) arc (70:290: 0.15cm and  0.4cm);
\draw[->,>=latex] (axis cs:70,80) -- (axis cs:83,84);
\draw[->,>=latex] (axis cs:70,80) -- (axis cs:95,43);
\draw[->,>=latex] (axis cs:70,80) -- (axis cs:78,49);

\node[align=center,fill=white,inner sep=3pt] (NoPC) at (axis cs: 50,10) {$D=125$ m};
\draw (axis cs: 50,59) arc (70:290: 0.15cm and  0.4cm);
\draw[->,>=latex] (axis cs:47,13) -- (axis cs:35,23);
\draw[->,>=latex] (axis cs:47,13) -- (axis cs:49,49);
\draw[->,>=latex] (axis cs:47,13) -- (axis cs:59,30);


\end{axis}

\end{tikzpicture} 
	\caption{Minimum Effective rate of non-FL UEs for different values of number of BS antennas. Here $L=K=5$.}
	\label{dataVSantennas} 
\end{figure}

\begin{figure}[t!]
	\centering
	\pgfplotsset{width=0.6\columnwidth,height =0.48\columnwidth, compat=1.6}
	\setlength{\baselineskip}{0pt}
\begin{tikzpicture}
\begin{axis}[
xtick={2,4,...,8},
ymin=0,ymax=110,xmin=2,xmax=8,
grid=both,
minor tick num = 1,
grid style={mygray},
legend style={anchor=north east,draw=black,fill=white,legend cell align=left,inner sep=1pt,row sep = -3pt,at={(0.38,0.275)}},
xlabel={Number of FL UEs ($L$)},
ylabel={Min. Effect. Rate of Non-FL UEs (Mbps)},
legend entries={BL1 Scheme, BL2 Scheme, Algorithm $1$, Algorithm $2$}
]
\addlegendimage{mark=x,line width=\lw, draw=mygreen}
\addlegendimage{mark=o,line width=\lw, draw=myred}
\addlegendimage{mark=square,line width=\lw, draw=myblue}
\addlegendimage{mark=*,line width=\lw, draw=myblack}

\addplot[mark=o,dashed,line width=\lw, draw=myred] table [y=EPA1, x=L,col sep = comma]{FLUEsJ3.csv};
\addplot[mark=x,dashed,line width=\lw, draw=mygreen] table [y=BL1, x=L,col sep = comma]{FLUEsJ3.csv};
\addplot[mark=square,dashed,line width=\lw, draw=myblue] table [y=HD1, x=L,col sep = comma]{FLUEsJ3.csv};
\addplot[mark=*,dashed,line width=\lw, draw=myblack] table [y=FD1, x=L,col sep = comma]{FLUEsJ3.csv};
\addplot[mark=o,line width=\lw, draw=myred] table [y=EPA2, x=L,col sep = comma]{FLUEsJ3.csv};
\addplot[mark=x,line width=\lw, draw=mygreen] table [y=BL2, x=L,col sep = comma]{FLUEsJ3.csv};
\addplot[mark=square,line width=\lw, draw=myblue] table [y=HD2, x=L,col sep = comma]{FLUEsJ3.csv};
\addplot[mark=*,line width=\lw, draw=myblack] table [y=FD2, x=L,col sep = comma]{FLUEsJ3.csv};

%
%

\end{axis}

\begin{axis}[
ymin=0,ymax=110,xmin=2,xmax=8,
legend style={anchor=north east,draw=black,fill=white,legend cell align=left,inner sep=1pt,row sep = -3pt,at={(0.68,0.16)}},
legend entries={$M=50$, $M=100$}
]
\addlegendimage{dashed,line width=\lw, draw=myblack}
\addlegendimage{line width=\lw, draw=myblack}
\end{axis}

\end{tikzpicture} 
	\caption{Minimum Effective rate of non-FL UEs for different values of number of FL UEs. Here $K=5$, and $M=50$.}
	\label{datavsFLUEs} 
\end{figure}
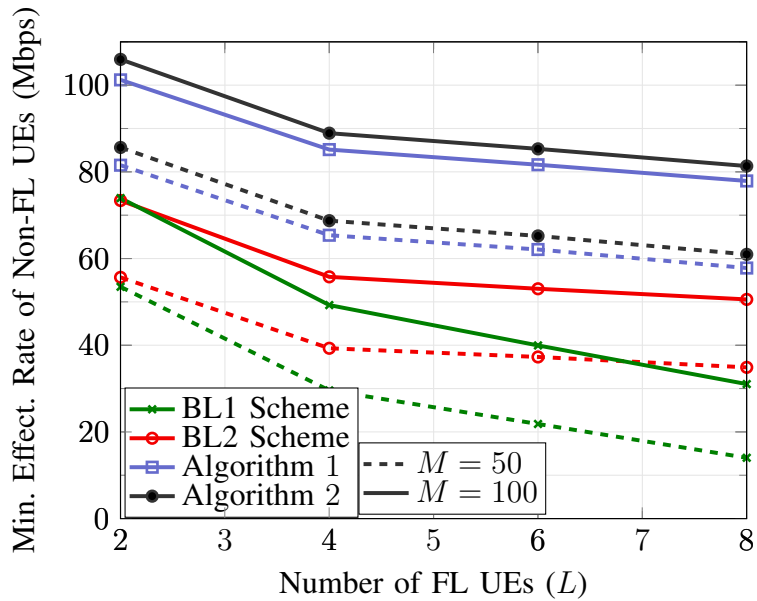

Next, in Figs.~\ref{dataVSantennas} and~\ref{datavsFLUEs}, we compare the minimum effective rate of the non-FL UEs by the two proposed schemes and the two considered baseline schemes. As seen clearly, both proposed schemes offer a better performance than the baseline counterparts. 
The figures not only demonstrate the significant advantage of a joint allocation of power and computing frequency over the heuristic scheme \textbf{BL2}, but also show the benefit of using massive MIMO. Thanks to massive MIMO technology, the data rate of each non-FL UE increases when the number of antennas increases, which then leads to a significant increase in the minimum effective data rates. 

Moreover, Figs.~\ref{dataVSantennas} and~\ref{datavsFLUEs} also confirm that in each frequency band used for each group, serving all the UEs simultaneously is better than serving them using the EPA approach. Specifically, the proposed approaches outperform \textbf{BL2} in almost every case. The gap between the proposed schemes and \textbf{BL1} is even bigger when the number of FL UEs increases.
This is because the effective rate of non-FL UE $k$ can be considered as the weighted rate $R_k\triangleq \frac{R_{1,k}t_d+R_{2,k}t_C+R_{3,k}t_u}{t_d+t_C+t_u}$, where $t_d, t_C, t_u$ are the weights associated with $R_{1,k}$, $R_{2,k}$, and $R_{3,k}$, respectively. Here, $R_{2,k}$ is the dominant rate because in Step (S2), all the non-FL UEs are served simultaneously without interference from FL UEs. In \textbf{BL1}, $R_{u,\ell}$ is very small due to its prelog factor $\frac{1}{L+K}$, which leads to a large $t_u$. When the weight $t_u$ becomes dominant compared to $t_d$ and $t_C$, $R_k$ of \textbf{BL1} is close to $R_{u,\ell}$ which is much lower than $R_k$ of the proposed schemes. As $L$ increases, $R_{u,\ell}$ decreases further and hence, $R_k$ also decreases.

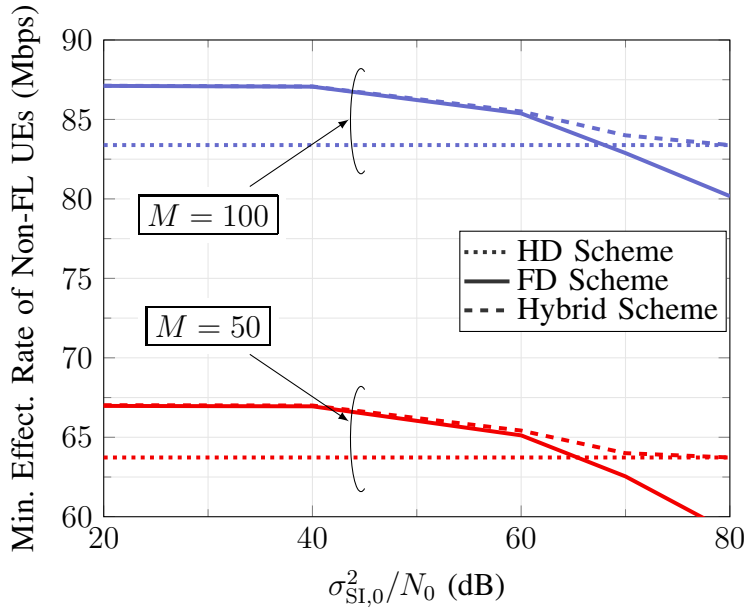
\begin{figure}[t!]
\centering
\pgfplotsset{width=0.6\columnwidth,height =0.48\columnwidth, compat=1.6}
\setlength{\baselineskip}{0pt}
\begin{tikzpicture}
\begin{axis}[
xtick={20,40,60,80},
ymin=60,ymax=90,xmin=20,xmax=80,
grid=both,
minor tick num = 1,
grid style={mygray},
legend style={anchor=north east,draw=black,fill=white,legend cell align=left,inner sep=1pt,row sep = -3pt,at={(0.995,0.6)}},
xlabel={$\sigma_{\mathrm{SI},0}^2/N_0$ (dB)},
ylabel={Min. Effect. Rate of Non-FL UEs (Mbps)},
legend entries={HD Scheme, FD Scheme, Hybrid Scheme}
]
\addlegendimage{dotted,myblack,line width=\lw}
\addlegendimage{myblack,line width=\lw}
\addlegendimage{dashed,myblack,line width=\lw}

\addplot[dotted,mark=none, line width=\lw, draw=myred] table [y=HD11, x=SI,col sep = comma]{HDvsFDJ5.csv};
\addplot[dotted,mark=none, line width=\lw, draw=myblue] table [y=HD22, x=SI,col sep = comma]{HDvsFDJ5.csv};
\addplot[mark=none, line width=\lw, draw=myred] table [y=FD11, x=SI,col sep = comma]{HDvsFDJ5.csv};
\addplot[mark=none, line width=\lw, draw=myblue] table [y=FD22, x=SI,col sep = comma]{HDvsFDJ5.csv};
\addplot[dashed,mark=none, line width=\lw, draw=myred] table [y=HDFD11, x=SI,col sep = comma]{HDvsFDJ5.csv};
\addplot[dashed,mark=none, line width=\lw, draw=myblue] table [y=HDFD22, x=SI,col sep = comma]{HDvsFDJ5.csv};

\node[align=center,fill=white,inner sep=3pt] (NoPC) at (axis cs: 30,79) {\boxed{M=100}};
\draw (axis cs: 45, 88) arc (70:290: 0.14cm and  0.7cm);
\draw[->,>=latex] (NoPC) -- (axis cs:43.5,85);
\node[align=center,fill=white,inner sep=3pt] (NoPC) at (axis cs: 30,72) {\boxed{M=50}};
\draw (axis cs: 45, 68) arc (70:290: 0.14cm and  0.7cm);
\draw[->,>=latex] (NoPC) -- (axis cs:43.5,66);
%



\end{axis}
\end{tikzpicture} 
\caption{Minimum Effective Rate of non-FL UEs for different values of $\sigma^2_{\text{SI},0}/N_0$ in dB.}
\label{HDvsFD}
\end{figure}

We now investigate the effect of high SI on the performance of the FD-based solution to understand when the FD-based algorithm is superior to the HD counterpart. For this purpose, the minimum effective rate of non-FL UEs is plotted in Fig. \ref{HDvsFD} for different values of $\sigma^2_{\text{SI},0}/N_0$. We also introduce a hybrid scheme which selects the approach that has the better objective among the two. For low values of SI (i.e., upto $65$ dB), the FD-based approach performs better,  which is expected and thus, the hybrid scheme is the same as the FD-based scheme. On the other hand, for large values of SI (i.e., beyond $65$ dB), the effectiveness of the FD-based approach starts to decrease due to the increased SI between the FL and non-FL groups. Especially, the HD-based scheme outperforms the FD-based approach when the SI is around than $80$ dB. Thus, the hybrid scheme is equal to the HD-based scheme for very large SI as can be seen clearly in Fig. \ref{HDvsFD}. 

\begin{figure}[t!]
\subfloat[Min. Effect. Rate of Non-FL UEs VS $S_d=S_u$]{\label{adv1}\pgfplotsset{width=0.48\columnwidth,height =0.45\columnwidth, compat=1.6}
\setlength{\baselineskip}{0pt}
\begin{tikzpicture}
\begin{axis}[
xtick={8,16,24,32,40},
ymin=55,ymax=90,xmin=8,xmax=40,
grid=both,
minor tick num = 1,
grid style={mygray},
legend style={anchor=north east,draw=black,fill=white,legend cell align=left,inner sep=1pt,row sep = -3pt,at={(0.99,0.55)}},
xlabel={$S_d=S_u$ (Mb)},
ylabel={Min. Effect. Rate of Non-FL UEs (Mbps)},
ylabel near ticks,
legend entries={Algorithm $1$, Algorithm $2$}
]
\addlegendimage{dashed,mark=none,myblack,line width=\lw}
\addlegendimage{mark=none,myblack,line width=\lw}

\addplot[dashed,mark=triangle, line width=\lw, draw=myred] table [y=HD11, x=SdSu,col sep = comma]{advantageJ2.csv};
\addplot[dashed,mark=square, line width=\lw, draw=myred] table [y=HD22, x=SdSu,col sep = comma]{advantageJ2.csv};
\addplot[mark=triangle, line width=\lw, draw=myblue] table [y=FD11, x=SdSu,col sep = comma]{advantageJ2.csv};
\addplot[mark=square, line width=\lw, draw=myblue] table [y=FD22, x=SdSu,col sep = comma]{advantageJ2.csv};

\node[align=center,fill=white,inner sep=3pt] (NoPC) at (axis cs: 15,59.5) {$M=50$};
\draw (axis cs: 21,67.5) arc (70:290: 0.14cm and  0.6cm);
\draw[->,>=latex] (NoPC) -- (axis cs:20,65);
\node[align=center,fill=white,inner sep=3pt] (NoPC) at (axis cs: 15,79) {$M=100$};
\draw (axis cs: 21,87.5) arc (70:290: 0.14cm and  0.6cm);
\draw[->,>=latex] (NoPC) -- (axis cs:20,84.5);

\end{axis}
\end{tikzpicture}} 
\centering 
\subfloat[Percentage advantage VS $S_d=S_u$]{\label{adv2}\pgfplotsset{width=0.48\columnwidth,height =0.45\columnwidth, compat=1.6}
\setlength{\baselineskip}{0pt}
\begin{tikzpicture}
\begin{axis}[
xtick={8,16,24,32,40},
ymin=3,ymax=13,xmin=8,xmax=40,
grid=both,
minor tick num = 1,
grid style={mygray},
legend style={anchor=north east,draw=black,fill=white,legend cell align=left,inner sep=1pt,row sep = -3pt,at={(0.42,0.95)}},
xlabel={$S_d=S_u$ (Mb)},
ylabel={$\mu$},
ylabel near ticks,
legend entries={$M=50$, $M=100$}
]
\addlegendimage{mark=triangle, line width=\lw, draw=myred}
\addlegendimage{mark=square, line width=\lw, draw=myblue}

\addplot[mark=triangle, line width=\lw, draw=myred] table [y=adv1, x=SdSu,col sep = comma]{advantageJ2.csv};
\addplot[mark=square, line width=\lw, draw=myblue] table [y=adv2, x=SdSu,col sep = comma]{advantageJ2.csv};


\end{axis}
\end{tikzpicture}} 
\caption{Minimum effective rate of non-FL UEs for different values of $S_d$ and $S_u$. Here $L=K=5$. }
\label{advantage}
\end{figure}
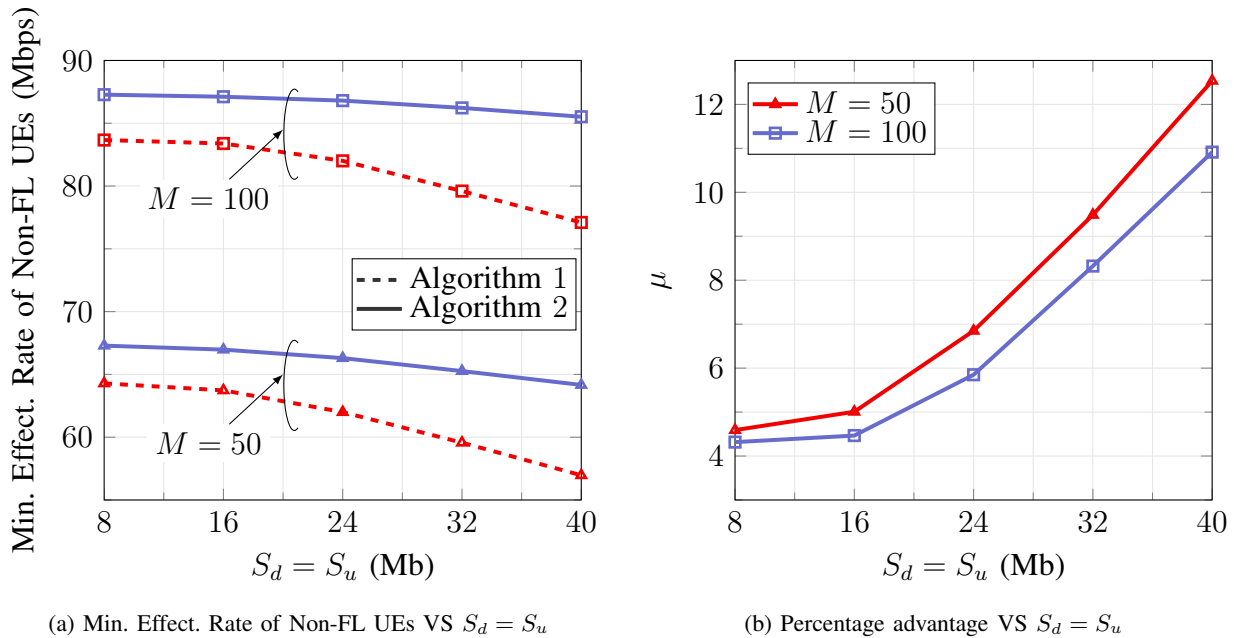

In the final numerical experiment, we plot the minimum effective rate of non-FL UEs against the values of $S_d$ and $S_u$ in Fig. \ref{advantage} to compare the performance of both proposed algorithms for different data sizes. We introduce a parameter $\mu\triangleq\frac{R_k^{\text{FD}}-R_k^{\text{HD}}}{R_k^{\text{HD}}}\times100$, which defines the gain in   percentage of the FD-based solution over the HD-based solution.
From  Fig. \ref{advantage}, we can observe that as the data size increases, the performance of FD-based scheme decreases very slowly compared to the HD-based scheme. As a result, the gains in percentage of the FD-based scheme over the HD-based scheme, $\mu$, increases with the data size. Thus, we can conclude that for problems with large sizes of FL model updates, the FD-based scheme should be preferred over the HD-based scheme.

\section{Conclusion}

\label{sec:con} We have presented two communication schemes that can support both the FL
and non-FL UEs using a massive MIMO technology which has not been considered previously. In particular, we have defined and maximize the effective rate of downlink
non-FL UEs in presence of a QoS latency constraint on FL UEs. We have also presented the HD and FD based solutions to the considered problem, specifically during the uplink of each FL iteration where the uplink FL UEs and downlink non-FL UEs receive their corresponding data simultaneously. In the downlink of each FL iteration, both FL and non-FL UEs continue to be served in the same time-frequency resource. 
The simulation results have showed that
the proposed HD-based and FD-based schemes outperform  the considered baseline schemes in all considered scenarios. It has also been shown that the FD-based scheme is superior to the HD-based scheme for the SI of upto $70$ dB. The FD-based scheme is also more beneficial than the HD-based scheme in terms of th effective rate achieved by the non-FL UEs in the cases of large sizes of the model updates.

\appendices{}

\section{Achievable Rates for FD in Step (S3)}
\label{sec:FDrates}

\textbf{Uplink transmission of FL UEs}: In this appendix, we simplify \eqref{eq:SINR_FL_S3-1}. Particularly,
we need to find two variance terms: $\var\{\uu_{u,\ell}^{H}\E_{u}\D_{\ETA_{u}}^{1/2}\s_{u}\}$
and $\var\{\uu_{u,\ell}^{H}\G_{\mathrm{SI}}\U_{3}\D_{\ZETA_{3}}^{1/2}\s_{u}\}$.
For the first term, we know that ${\Z}_{u}$ is independent of
${\E}_{u}$. Thus, we can state that 
\begin{align}
\var\{\uu_{u,\ell}^{H}\E_{u}\D_{\ETA_{u}}^{1/2}\s_{u}\} & =\sum_{i\in\LL}({\beta}_{i}-{\sigma}_{u,i}^{2}){\eta}_{u,i}.
\end{align}
\begin{align}
\var\{\uu_{u,\ell}^{H}\G_{\mathrm{SI}}\U_{3}\D_{\ZETA_{3}}^{1/2}\s_{u}\} & =\EEE\{|\uu_{u,\ell}^{H}\G_{\mathrm{SI}}\U_{3}\D_{\ZETA_{3}}^{1/2}\s_{u}|^{2}\}\nonumber \\
 & =\EEE\{\uu_{u,\ell}^{H}\G_{\mathrm{SI}}\U_{3}\D_{\ZETA_{3}}^{1/2}\s_{u}\s_{u}^{H}\D_{\ZETA_{3}}^{1/2}\U_{3}^{H}\G_{\mathrm{SI}}^{H}\uu_{u,\ell}\}\nonumber \\
 & =\EEE\{\uu_{u,\ell}^{H}\G_{\mathrm{SI}}\U_{3}\D_{\ZETA_{3}}\U_{3}^{H}\G_{\mathrm{SI}}^{H}\uu_{u,\ell}\}\nonumber \\
 & =(\sum_{i\in\K}\zeta_{3,i})\EEE\{\uu_{u,\ell}^{H}\G_{\mathrm{SI}}\U_{3}\U_{3}^{H}\G_{\mathrm{SI}}^{H}\uu_{u,\ell}\}\nonumber \\
 & =(\sum_{i\in\K}\zeta_{3,i})\EEE\{\uu_{u,\ell}^{H}\G_{\mathrm{SI}}\G_{\mathrm{SI}}^{H}\uu_{u,\ell}\}.
\end{align}
Using law of large numbers, $\EEE\{\G_{\mathrm{SI}}\G_{\mathrm{SI}}^{H}\}\approx M\beta_{\text{SI}}\sigma_{\mathrm{SI},0}^{2}$. Therefore, the above equation can be approximated as
\begin{align}
\var\{\uu_{u,\ell}^{H}\E_{u}\D_{\ETA_{u}}^{1/2}\s_{u}\} & \approx M\beta_{\text{SI}}\sigma_{\mathrm{SI},0}^{2}(\sum_{i\in\K}\zeta_{3,i})\EEE\{\uu_{u,\ell}^{H}\uu_{u,\ell}\}\nonumber \\
& =M\beta_{\text{SI}}\sigma_{\mathrm{SI},0}^{2}\sum_{i\in\K}\zeta_{3,i}.
\end{align}
\section{Achievable Rates and Data Received for Baseline 1 (BL1) Scheme}
\label{sec:BL1rates}

For the FD communication in Step (S3), FL UEs observe SI from non-FL UEs, while non-FL UEs observe IGI from FL UEs. Assume the FDMA approach is used in this scheme. The bandwidth resources for FL UEs and non-FL UEs are divided by $(L+K)$. Hence, the uplink rate is expressed as 
\begin{equation}
R_{u}^\text{FDMA}(\ETA_{u},\ZETA_{3})\triangleq\min_{\ell\in\LL}\frac{\tau_{c}-\tau_{u,p}}{(L+K)\tau_{c}}B\log_{2}\big(1+\text{SINR}^\text{FDMA}_{u,\ell}(\ETA_{u},\ZETA_{3})\big),
\end{equation}
where $\tau_{u,p}=1$ and $\text{SINR}^\text{FDMA}_{u,\ell}(\ETA_{u})$ is given by
\begin{align}
\text{SINR}^\text{FDMA}_{u,\ell}(\ETA_{u},\ZETA_{3})
=\frac{\rho_{u}\eta_{u,\ell}M\sigma_{u,\ell}^{2}}{1+\rho_{u}\beta_{\ell}\eta_{u,\ell}}.
\end{align}
Then, the time for each FL UE to complete its uplink transmission is the same as
\begin{align}
    t_u^{\text{FDMA}}(\ETA_{u}) = \frac{S_u}{R_{u}^\text{FDMA}(\ETA_{u})}.
\end{align}

Similarly for non-FL UEs, the downlink rate can be provided as 
\begin{equation}
R_{3,k}^\text{FDMA}(\ETA_{u},\ZETA_{3})=\frac{\tau_{c}-\tau_{3,p}}{(L+K)\tau_{c}}B\log_{2}\big(1+\text{SINR}_{3,k}^{\text{FDMA}}(\ETA_{u},\ZETA_{3})\big),
\end{equation}
where $\tau_{3,p}=1$ and $\text{SINR}^\text{FDMA}_{3,k}(\ETA_{u},\ZETA_{3})$ is calculated
as 
\begin{align}
\text{SINR}^\text{FDMA}_{3,k}(\ETA_{u},\ZETA_{3}) & 
=\frac{\rho_{d}\zeta_{3,k}M\sigma_{3,k}^{2}}{1+\rho_{d}\bar{\beta}_{k}\zeta_{3,k}}.
\end{align}
The transmission time from each FL UE to the BS and the
amount of downlink data received at all non-FL UE $k,\forall k\in\K$
are given by \eqref{eq:TimeFLStep3} and \eqref{eq:DataNonFLStep3},
respectively.
The optimization problem in this scheme is
\begin{subequations}
\label{Pmain:BL1:FD} 
\begin{align}
\!\!\!\!\!\!\!\!\underset{\x^\text{FDMA}}{\max}\,\, & \frac{\min_{k\in\K}\big(D_{1,k}(\ETA_{d},\ZETA_{1})\!+\!D_{2,k}(f,\ZETA_{2})\!+\!D_{3,k}^{\text{FDMA}}(\ETA_{u},\ZETA_{3})\big)}{t_{d}(\ETA_{d},\ZETA_{1})+t_{C}(f)+t_{u}^{\text{ FMDA}}(\ETA_{u})}\\
\mathrm{s.t.}\,\, & \eqref{powerdupperbound},\eqref{powerdupperbound-1},\eqref{poweruupperbound},\eqref{powerdupperbound-2},\eqref{S1powerUbound},\eqref{fbound},\nonumber\\
 & t_{d}(\ETA_{d},\ZETA_{1})+t_{C}(f)+ t_{u}^{\text{FDMA}}(\ETA_{u}) \leq t_{\text{QoS}}.
\end{align}
\end{subequations}

\section{Expressions of lower and upper bounds of rate functions 
\label{sec:LBrates} }
From \eqref{eq:logapproxLB}, the concave lower bounds of of $R_{d,\ell}(\ETA_{d},\ZETA_{1})$, $R_{u,\ell}^{\text{HD}}(\ETA_{u})$,
$R_{1,k}(\ETA_{d},\ZETA_{1})$, $R_{2,k}(\ZETA_{2})$, and $R_{3,k}^{\text{HD}}(\ZETA_{3})$
are found as
{\small
\begin{subequations}\label{eq:rate:HD:LB}
\begin{align}
\tilde{R}_{d,\ell}(\ETA_{d},\ZETA_{1})& \!=\!\frac{\tau_{c}\!-\!\tau_{d,p}}{\tau_{c}\log2}B\Big[\log\big(1\!+\!\frac{\PSI_{d,\ell}^{(n)}}{\THETA_{d,\ell}^{(n)}}\big)\!+\!\frac{2\PSI_{d,\ell}^{(n)}}{\PSI_{d,\ell}^{(n)}\!+\!\THETA_{d,\ell}^{(n)}}\!-\!\frac{(\PSI_{d,\ell}^{(n)})^{2}}{(\PSI_{d,\ell}^{(n)}\!+\!\THETA_{d,\ell}^{(n)})\PSI_{d,\ell}}\!-\!\frac{\PSI_{d,\ell}^{(n)}\THETA_{d,\ell}}{(\PSI_{d,\ell}^{(n)}\!+\!\THETA_{d,\ell}^{(n)})\THETA_{d,\ell}^{(n)}}\Big] 
\label{Rd_tilde_apprx},\\
\tilde{R}_{u,\ell}^{\text{HD}}(\!\ETA_{u}\!)
& \!\!=\!\frac{\tau_{c}\!-\!\tau_{u,p}}{\tau_{c}\log2}\frac{B}{2}\!\Big[\!\log\big(1\!\!+\!\frac{\PSI_{u,\ell}^{(n)}}{\THETA_{u,\ell}^{\text{HD}(n)}}\big)\!\!+\!\frac{2\PSI_{u,\ell}^{(n)}}{\PSI_{u,\ell}^{(n)}\!\!+\!\THETA_{u,\ell}^{\text{HD}(n)}}\!-\!\frac{(\PSI_{u,\ell}^{(n)})^{2}}{(\PSI_{u,\ell}^{(n)}\!\!+\!\THETA_{u,\ell}^{\text{HD}(n)}\!)\PSI_{u,\ell}}\!-\!\frac{\PSI_{u,\ell}^{(n)}\THETA_{u,\ell}^{\text{HD}}}{(\PSI_{u,\ell}^{(n)}\!\!+\!\THETA_{u,\ell}^{\text{HD}(n)}\!)\THETA_{u,\ell}^{\text{HD}(n)}}\!\Big],
\label{Ru_tilde_approx}\\
\tilde{R}_{1,k}(\ETA_{d},\ZETA_{1}) & \!=\!\frac{\tau_{c}\!-\!\tau_{1,p}}{\tau_{c}\log2}B\Big[\log\!\big(1\!+\!\frac{\PSI_{1,k}^{(n)}}{\THETA_{1,k}^{(n)}}\big)\!+\!\frac{2\PSI_{1,k}^{(n)}}{\PSI_{1,k}^{(n)}\!+\!\THETA_{1,k}^{(n)}}\!-\!\frac{(\PSI_{1,k}^{(n)})^{2}}{(\PSI_{1,k}^{(n)}\!+\!\THETA_{1,k}^{(n)})\PSI_{1,k}}\!-\!\frac{\PSI_{1,k}^{(n)}\THETA_{1,k}}{(\PSI_{1,k}^{(n)}\!+\!\THETA_{1,k}^{(n)})\THETA_{1,k}^{(n)}}\Big], 
\label{R1k_tilde_approx}\\
\tilde{R}_{2,k}(\ZETA_{2}) & \!=\!\frac{\tau_{c}\!-\!\tau_{2,p}}{\tau_{c}\log2}B\Big[\log\!\big(1\!+\!\frac{\PSI_{2,k}^{(n)}}{\THETA_{2,k}^{(n)}}\big)\!+\!\frac{2\PSI_{2,k}^{(n)}}{\PSI_{2,k}^{(n)}\!+\!\THETA_{2,k}^{(n)}}\!-\!\frac{(\PSI_{2,k}^{(n)})^{2}}{(\PSI_{2,k}^{(n)}\!+\!\THETA_{2,k}^{(n)})\PSI_{2,k}}\!-\!\frac{\PSI_{2,k}^{(n)}\THETA_{2,k}}{(\PSI_{2,k}^{(n)}\!+\!\THETA_{2,k}^{(n)})\THETA_{2,k}^{(n)}}\Big], 
\label{R2k_tilde_approx}\\
\tilde{R}_{3,k}^\text{HD}(\!\ETA_{u},\!\ZETA_{3}\!) & \!\!=\!\!\frac{\tau_{c}\!-\!\tau_{3,p}}{\tau_{c}\log2}\frac{B}{2}\!\Big[\!\log\!\big(\!1\!\!+\!\frac{\PSI_{3,k}^{(n)}}{\THETA_{3,k}^{\text{HD}(n)}}\!\big)\!\!+\!\!\frac{2\PSI_{3,k}^{(n)}}{\PSI_{3,k}^{(n)}\!\!+\!\!\THETA_{3,k}^{\text{HD}(n)}}\!-\!\frac{(\PSI_{3,k}^{(n)})^{2}}{(\PSI_{3,k}^{(n)}\!\!+\!\!\THETA_{3,k}^{\text{HD}(n)}\!)\!\PSI_{3,k}}\!-\!\frac{\PSI_{3,k}^{(n)}\THETA_{3,k}^{\text{HD}}}{(\PSI_{3,k}^{(n)}\!\!+\!\THETA_{3,k}^{\text{HD}(n)}\!)\!\THETA_{3,k}^{\text{HD}(n)}}\Big], 
\label{R3k_tilde_approx}
\end{align}
\end{subequations}
}
where 
$\PSI_{d,\ell}\!=\!\rho_{d}(M\!-\!L\!-\!K)\sigma_{d,\ell}^{2}\eta_{d,\ell},
\PSI_{d,\ell}^{(n)}\!=\!\rho_{d}(M\!-\!L\!-\!K)\sigma_{d,\ell}^{2}\eta_{d,\ell}^{(n)},
\THETA_{d,\ell}\!=\!1\!+\!\rho_{d}(\beta_{\ell}\!-\!\sigma_{d,\ell}^{2})\sum_{i\in\LL}\eta_{d,i}\!+\!\rho_{d}\beta_{\ell}\sum_{k\in\K}\zeta_{1,k},
\THETA_{d,\ell}^{(n)}\!=\!1\!+\!\rho_{d}(\beta_{\ell}\!-\!\sigma_{d,\ell}^{2})\sum_{i\in\LL}\eta_{d,i}^{(n)}\!+\!\rho_{d}\beta_{\ell}\sum_{k\in\K}\zeta_{1,k}^{(n)},
\PSI_{u,\ell}\!=\!\rho_{u}(M\!-\!L)\sigma_{u,\ell}^{2}\eta_{u,\ell},
\PSI_{u,\ell}^{(n)}\!=\!\rho_{u}(M\!-\!L)\sigma_{u,\ell}^{2}\eta_{u,\ell}^{(n)},
\THETA_{u,\ell}^{\text{HD}}\!=\!1\!+\!\rho_{u}\sum_{i\in\LL}(\beta_{i}\!-\!\sigma_{u,i}^{2})\eta_{u,i},
\THETA_{u,\ell}^{\text{HD}(n)}\!=\!1\!+\!\rho_{u}\sum_{i\in\LL}(\beta_{i}\!-\!\sigma_{u,i}^{2})\eta_{u,i}^{(n)},
\PSI_{1,k}=\rho_{d}(M\!-\!L\!-\!K)\sigma_{1,k}^{2}\zeta_{1,k},
\PSI_{1,k}^{(n)}=\rho_{d}(M\!-\!L\!-\!K)\sigma_{1,k}^{2}\zeta_{1,k}^{(n)},
\THETA_{1,k}=1\!+\!\rho_{d}(\bar{\beta}_{k}\!-\!\sigma_{1,k}^{2})\sum_{i\in\K}\zeta_{1,i}\!+\!\rho_{d}\bar{\beta}_{k}\sum_{\ell\in\LL}\eta_{d,\ell},
\THETA_{1,k}^{(n)}=1\!+\!\rho_{d}
(\bar{\beta}_{k}\!-\!\sigma_{1,k}^{2})\sum_{i\in\K}\zeta_{1,i}^{(n)}\!+\!\rho_{d}\bar{\beta}_{k}\sum_{\ell\in\LL}\eta_{d,\ell}^{(n)};\PSI_{2,k}=\rho_{d}(M\!-\! K)\sigma_{1,k}^{2}\zeta_{2,k},\PSI_{2,k}^{(n)}=\rho_{d}(M\!-\!K)\sigma_{k}^{2}\zeta_{2,k}^{(n)},
\THETA_{2,k}=1\!+\!\rho_{d}(\bar{\beta}_{k}\!-\!\sigma_{2,k}^{2})\sum_{i\in\K}\zeta_{2,i},
\THETA_{2,k}^{(n)}=1\!+\!\rho_{d}(\bar{\beta}_{k}\!-\!\sigma_{2,k}^{2})\sum_{i\in\K}\zeta_{2,i}^{(n)},
\PSI_{3,k}=\rho_{d}(M\!-\!K)\sigma_{3,k}^{2}\zeta_{3,k},
\PSI_{3,k}^{(n)}=\rho_{d}(M\!-\!K)\sigma_{3,k}^{2}\zeta_{3,k}^{(n)},
\THETA_{3,k}^{\text{HD}}=1\!+\!\rho_{d}(\bar{\beta}_{k}\!-\!\sigma_{3,k}^{2})\sum_{i\in\K}\zeta_{3,i},
\THETA_{3,k}^{\text{HD}(n)}\!=\!1\!+\!\rho_{d}(\bar{\beta}_{k}\!-\!\sigma_{3,k}^{2}) \sum_{i\in\K}\zeta_{3,i}^{(n)}$.

From \eqref{eq:logapproxUB} convex upper bounds of ${R}_{d,\ell}(\ETA_{d},\ZETA_{1})$ and ${R}_{u,\ell}^\text{HD}(\ETA_{u})$ are given by
{\small
\begin{subequations}\label{eq:rate:HD:UB}
\begin{align}
\hat{R}_{d,\ell}(\ETA_{d},\ZETA_{1}) & =\frac{\tau_{c}\!-\!\tau_{d,p}}{\tau_{c}\log2}B\Big[\log\big(1+\frac{\PSI_{d,\ell}^{(n)}}{\THETA_{d,\ell}^{(n)}}\big)+\frac{\THETA_{d,\ell}^{(n)}}{\PSI_{d,\ell}^{(n)}+\THETA_{d,\ell}^{(n)}}-\frac{(\PSI_{d,\ell})^{2}+(\PSI_{d,\ell}^{(n)})^{2}}{2\times\PSI_{d,\ell}^{(n)}\times\THETA_{d,\ell}}-\frac{\PSI_{d,\ell}^{(n)}}{\THETA_{d,\ell}^{(n)}}\Big],\\
\hat{R}_{u,\ell}^\text{HD}(\ETA_{u}) & =\frac{\tau_{c}\!-\!\tau_{u,p}}{\tau_{c}\log2}\frac{B}{2}\Big[\log\big(1+\frac{\PSI_{u,\ell}^{(n)}}{\THETA_{u,\ell}^{\text{HD}(n)}}\big)+\frac{\THETA_{u,\ell}^{\text{HD}(n)}}{\PSI_{u,\ell}^{(n)}+\THETA_{u,\ell}^{\text{HD}(n)}}-\frac{(\PSI_{u,\ell})^{2}+(\PSI_{u,\ell}^{(n)})^{2}}{2\PSI_{u,\ell}^{(n)}\THETA_{u,\ell}^\text{HD}}-\frac{\PSI_{u,\ell}^{(n)}}{\THETA_{u,\ell}^{\text{HD}(n)}}\Big].
\end{align}
\end{subequations}
}

Similarly, the concave lower bounds of $R_{u,\ell}^{\text{FD}}(\ETA_{u},\ZETA_{3})$ and $R_{3,k}^{\text{FD}}(\ETA_{u},\ZETA_{3})$ for the FD scheme are written as
{\small
\begin{subequations}\label{eq:rate:FD:LB}
\begin{align}
\tilde{R}_{u,\ell}^{\text{FD}}(\!\ETA_{u}\!)
& \!\!=\!\frac{\tau_{c}\!-\!\tau_{u,p}}{\tau_{c}\log2}\frac{B}{2}\!\Big[\!\log\big(1\!\!+\!\frac{\PSI_{u,\ell}^{(n)}}{\THETA_{u,\ell}^{\text{FD}(n)}}\big)\!\!+\!\frac{2\PSI_{u,\ell}^{(n)}}{\PSI_{u,\ell}^{(n)}\!\!+\!\THETA_{u,\ell}^{\text{FD}(n)}}\!-\!\frac{(\PSI_{u,\ell}^{(n)})^{2}}{(\PSI_{u,\ell}^{(n)}\!\!+\!\THETA_{u,\ell}^{\text{FD}(n)}\!)\PSI_{u,\ell}}\!-\!\frac{\PSI_{u,\ell}^{(n)}\THETA_{u,\ell}^{\text{FD}}}{(\PSI_{u,\ell}^{(n)}\!\!+\!\THETA_{u,\ell}^{\text{FD}(n)}\!)\THETA_{u,\ell}^{\text{FD}(n)}}\!\Big], 
\label{Ru_tilde_approx-1}\\
\tilde{R}_{3,k}^\text{FD}(\!\ETA_{u},\!\ZETA_{3}\!) & \!\!=\!\!\frac{\tau_{c}\!-\!\tau_{3,p}}{\tau_{c}\log2}\frac{B}{2}\!\Big[\!\log\!\big(\!1\!\!+\!\frac{\PSI_{3,k}^{(n)}}{\THETA_{3,k}^{\text{FD}(n)}}\!\big)\!\!+\!\!\frac{2\PSI_{3,k}^{(n)}}{\PSI_{3,k}^{(n)}\!\!+\!\!\THETA_{3,k}^{\text{FD}(n)}}\!-\!\frac{(\PSI_{3,k}^{(n)})^{2}}{(\PSI_{3,k}^{(n)}\!\!+\!\!\THETA_{3,k}^{\text{FD}(n)}\!)\!\PSI_{3,k}}\!-\!\frac{\PSI_{3,k}^{(n)}\THETA_{3,k}^{\text{FD}}}{(\PSI_{3,k}^{(n)}\!\!+\!\THETA_{3,k}^{\text{FD}(n)}\!)\!\THETA_{3,k}^{\text{FD}(n)}}\Big], 
\label{R3k_tilde_approx-1}
\end{align}
\end{subequations}}
and the convex upper bound of ${R}_{u,\ell}^\text{FD}(\ETA_{u})$ is given by
{\small
\begin{align}\label{eq:rate:FD:UB}
\hat{R}_{u,\ell}^\text{FD}(\ETA_{u}) & =\frac{\tau_{c}\!-\!\tau_{u,p}}{\tau_{c}\log2}\frac{B}{2}\Big[\log\big(1+\frac{\PSI_{u,\ell}^{(n)}}{\THETA_{u,\ell}^{\text{FD}(n)}}\big)+\frac{\THETA_{u,\ell}^{\text{FD}(n)}}{\PSI_{u,\ell}^{(n)}+\THETA_{u,\ell}^{\text{FD}(n)}}-\frac{(\PSI_{u,\ell})^{2}+(\PSI_{u,\ell}^{(n)})^{2}}{2\PSI_{u,\ell}^{(n)}\THETA_{u,\ell}^\text{FD}}-\frac{\PSI_{u,\ell}^{(n)}}{\THETA_{u,\ell}^{\text{FD}(n)}}\Big],
\end{align}
}
where $\THETA_{u,\ell}^{\text{FD}}=1+\rho_{u}\sum_{i\in\LL}(\beta_{i}-\sigma_{u,i}^{2})\eta_{u,i}+\rho_{d}M\beta_{\text{SI}}\sigma_{\mathrm{SI},0}^{2}\sum_{j\in\K}\zeta_{3,j},
\THETA_{u,\ell}^{\text{FD}(n)}=1+\rho_{u}\sum_{i\in\LL}(\beta_{i}-\sigma_{u,i}^{2})\eta_{u,i}^{(n)}+\rho_{d}M\beta_{\text{SI}}\sigma_{\mathrm{SI},0}^{2}\sum_{j\in\K}\zeta_{3,j}^{(n)},
\THETA_{3,k}^{\text{FD}}=1+\rho_{d}(\bar{\beta}_{k}-\sigma_{3,k}^{2})\sum_{i\in\K}\zeta_{3,i}+\rho_{u}\sum_{i\in\LL}\beta_{\mathrm{IGI},ki}\eta_{u,i},
\THETA_{3,k}^{\text{FD}(n)}=1+\rho_{d}(\bar{\beta}_{k}-\sigma_{3,k}^{2})\sum_{i\in\K}\zeta_{3,i}^{(n)}+\rho_{u}\sum_{i\in\LL}\beta_{\mathrm{IGI},ki}\eta_{u,i}^{(n)}.$

\bibliographystyle{IEEEtran}
\bibliography{IEEEabrv,newidea2021}

\end{document}